\documentclass[12pt]{article}

\usepackage{amsmath}
\usepackage[dvips]{graphicx}

\def\oa{\overline{a}}

\def\cN{\mathcal{N}}
\def\oPhi{\overline{\Phi}}
\def\ta{\tilde{a}}
\def\tchi{\tilde{\chi}}
\def\ttheta{\tilde{\theta}}
\def\oJ{\overline{J}}
\def\otheta{\overline{\theta}}
\def\ochi{\overline{\chi}}
\def\tchi{\tilde{\chi}}
\def\oa{\overline{a}}

\makeatletter
\@addtoreset{equation}{section}
\makeatother

\begin{document}

\begin{flushright}
May, 2008
\end{flushright}

\vspace{0.0cm}

\begin{center}

{\Large \bf 
Exact Lattice Supersymmetry at Large N 
}

\vspace{1.0cm}

{ Kazuhiro Nagata}\footnote{Email: knagata@indiana.edu}\\

\vspace{0.5cm}
{\it{ Department of Physics, Indiana University}}\\
{\it{ Bloomington, IN 47405, USA}}\\
\end{center}

\begin{abstract}
Employing a novel type of non-commutative product in 
the Dirac-K\"ahler twisted superspace on a lattice, 
we formulate a field theoretically rigid framework of 
extended supersymmetry on a lattice. 
As a first example of this treatment,
we calculate one-loop (in some cases any loops)
quantum corrections for 
a twisted Wess-Zumino model with $N\times N$ hermitian matrix superfields
on a two dimensional lattice.
The calculations are entirely given in a lattice superfield framework.
We report that the mass and the coupling constant are exactly protected
from the radiative corrections at non-zero lattice spacing
as far as the planar diagrams are concerned,
which implies the realization of exact lattice supersymmetry w.r.t. all 
the supercharges in the large-$N$ limit.

\end{abstract}

\section{Introduction}
\indent

There have been a variety of studies addressing 
the subject of supersymmetry (SUSY) on a lattice.
The literal purpose of lattice SUSY 
is to provide a well-defined constructive framework of 
SUSY formulations with which
one may extract the fully non-perturbative information 
either analytically or numerically. 
A successful formulation of lattice SUSY may be expected to 
provide a deeper understanding of the relationship
between bosons and fermions in a regularized framework.
In recent years, 
it has been recognized that
the so-called twisted SUSY formulations are playing particularly important roles
in this challenging subject
\cite{DKKN1,DKKN2,DKKN3,twist-lattsusy,Bruckmann,ADKS}.
It was also recently pointed out that 
amongst the other formulations
of lattice SUSY \cite{other-lattsusy1},
the deconstruction formulation \cite{deconstruction} 
is quite closely related to the 
twisted SUSY formulation on the lattice \cite{relation}. 
In spite of these developments, however,
realizing 
all the SUSY generators 
in a SUSY algebra being considered, 
say, $D=4\ \cN=1$ or $D=2\ \cN=2$ or
whatever \footnote{
In this paper, we shall represent the degree of extended SUSY by 
the calligraphic letter $\cN$ while
the size of the superfield matrices
by the letter $N$.},
exactly on a lattice,
has been considered as a very non-trivial task.
This is essentially because
in general only the finite subgroup of the entire
translational symmetry group in the continuum spacetime
can be survived at non-zero lattice spacings.

In the series of papers \cite{DKKN1,DKKN2,DKKN3},
we have proposed an extended SUSY formulation on a lattice
which keeps all the SUSY transformations exact 
at least at classical level. 
The formulation is based on the two crucial theoretical ingredients. 
The one is the notion of twisted SUSY. 
This is essentially originated from the 
intrinsic relation between twisted fermions and staggered fermions
or more appropriately Dirac-K\"ahler fermions \cite{KT}.
It has been pointed out that
the fermionic components in the lattice SUSY multiplet 
are realized as a staggered fermion,
and accordingly the spectrum doubling 
turns to have a physically relevant meaning
as an extended SUSY degrees of freedom \cite{DKKN1}.
The lattice fermions in this framework
have their own geometrical meaning.
They have one-to-one correspondence with all 
the simplices available on a lattice being considered,
namely, 0, 1 and 2-form on a two dimensional lattice
or 0, 1, 2, 3 and 4-form on a four dimensional lattice,
which is a natural consequence of the Dirac-K\"ahler fermion realization
on a lattice \cite{BJ}.        
The other crucial ingredient,
which is also tightly related to the Dirac-K\"ahler structure, 
is the ``mild" non-commutativity associated with the lattice supercharges.
It has been long recognized that the breakdown of Leibniz rule
is one of the major obstacles in realizing 
exact SUSY on a lattice.
The notion of ``mild" non-commutativity has been 
introduced in order to overcome this difficulty. 
The non-commutative nature of lattice supercharges
can be trace back to
the intrinsic non-commutative property 
of difference operators.
In \cite{DKKN1}, we have shown that if the non-commutativity
matching conditions, which we refer to as the Leibniz rule conditions 
on a lattice, are satisfied, 
then all the supercharges can be exactly realized on the lattice
at least at classical level.
This notion of ``mild" non-commutativity has been further applied to the
formulations of extended supersymmetric gauge theories 
by means of link supercharges and link component fields
\cite{DKKN2,DKKN3} and recently also applied to the formulation of
Chern-Simons theory on a three dimensional lattice \cite{Nagata-Wu}.

In spite of these developments, on the other hand, 
we have not addressed the 
quantum aspects of these formulations so far
since it has not been fully established how to deal with the above
``mild" non-commutativity in a quantum field theoretically rigid manner. 
In this paper, we attack this problem
and propose a field theoretically well-defined treatment
of the ``mild" non-commutative SUSY formulation on a lattice. 
In particular, we shall 
employ a novel type of non-commutative product 
in the Dirac-K\"ahler twisted superspace
which honestly accommodates the ``mild" non-commutativity
introduced in Ref. \cite{DKKN1}.
As a first example of this framework,
we shall calculate one-loop (in some cases any loops)
quantum corrections for 
a twisted Wess-Zumino model with $N\times N$ hermitian matrix superfields
on a two dimensional lattice.
The calculations are entirely given in the lattice superfield method.
We report that the mass and the coupling constant are exactly protected
from the quantum corrections as far as the planar diagrams are concerned,
while the SUSY spoiling non-planar contributions are suppressed
at least by $\mathcal{O}(a^{2}/N)$.
These features imply the realization of exact lattice supersymmetry w.r.t. all 
the supercharges in the 't Hooft large-$N$ limit.

This paper is organized as follows.
In Sec. 2, after introducing a novel type of non-commutative
product which honestly represents the ``mild" non-commutativity
which we have introduced in \cite{DKKN1},
we (re-)formulate the lattice counterpart of
extended SUSY algebra and see how to 
construct an extended SUSY invariant action
on a lattice.
In Sec. 3, as a simple and explicit example,
we introduce the Dirac-K\"ahler twisted $\cN=D=2$ SUSY algebra
and explicitly see how the lattice SUSY algebra is realized 
in accordance with the lattice Leibniz rule conditions. 
We then construct a $\cN=D=2$ twisted Wess-Zumino model on a lattice
and proceed to derive the superfield propagators and the three-point
vertex functions,
taking a full advantage of the lattice superfield method.
In Sec. 4, we extend the chiral and anti-chiral superfields 
introduced in Sec. 3
to $N\times N$ hermitian matrix valued superfields
in order to set up the $\cN=D=2$ twisted Wess-Zumino model
with global $U(N)$ symmetry.
We perturbatively investigate the possible radiative corrections to this model
by explicitly calculating the one-loop (in some cases any loops)
contributions to the kinetic term, the mass term and the interaction term
of the lattice action.
We then find out that 
the quantum corrections strictly protect the lattice SUSY
as far as the planar diagrams are concerned.
In particular, we shall see that 
apart from the overall wave function renormalization
the mass and the coupling constant
are strictly protected (at least to this order) 
from the radiative corrections in the planar diagrams.
These features imply the full realization of exact SUSY at
non-zero lattice spacing in the 't Hooft large-$N$ limit.
Sec. 5 summarizes the formulation 
with some discussions.

\section{Introducing a Non-Commutative Product} 
\indent

In this section, we first introduce a novel non-commutative
product in the Dirac-K\"ahler twisted superspace
which honestly represent the ``mild" non-commutativity
introduced in Ref. \cite{DKKN1}
and present how to formulate a lattice counterpart of SUSY algebra. 
The non-commutativity introduced in Ref \cite{DKKN1} 
in order to keep the lattice Leibniz rule 
is the one between 
the bosonic (discrete) coordinate $x_{\mu}$ and 
Grassmann coordinate $\theta_{A}$
in the twisted superspace $(x_{\mu},\theta_{A})$,
\begin{eqnarray}
\theta_{A}f(x) = (-1)^{|f|}f(x-2a_{A})\theta_{A},
\end{eqnarray}
or equivalently,
\begin{eqnarray}
[x_{\mu},\theta_{A}] &=& 2(a_{A})_{\mu}\theta_{A}, \label{NC1}
\end{eqnarray}
where the symbol $|f|$ takes value of $+1$ or $-1$ for
the bosonic or fermionic function $f$, respectively. 
$\theta_{A}$ is representing any fermionic coordinate 
in the Dirac-K\"ahler twisted superspace, namely, 
$\theta_{A}=(\theta, \theta_{\mu}, \tilde{\theta})$ 
with $\mu=1,2$ in $\cN=D=2$ \cite{DKKN1,DKKN2}, 
$\theta_{A}=(\theta, \theta_{\mu}, \overline{\theta}_{\mu},\overline{\theta})$
with $\mu=1\sim3$ in $\cN=4\ D=3$ \cite{DKKN3} and 
$\theta_{A}=(\theta, \theta_{\mu}, \theta_{\mu\nu}, \tilde{\theta}_{\mu},
\tilde{\theta})$ with $\mu,\nu=1\sim 4$ in $\cN=D=4$ \cite{DKKN2}.
The vector $a_{A}$ in the r.h.s. of (\ref{NC1})
governing
the non-commutativity between $x_{\mu}$ and $\theta_{A}$
is to be constrained at a later stage 
such that the resulting SUSY algebra 
be consistent with the non-commutativity matching conditions
which we refer to as 
the lattice Leibniz rule conditions.   
In this framework, superfields may generically be expanded
in terms of the non-commutative $\theta_{A}$'s
such as
\begin{eqnarray}
\Phi(x,\theta_{A}) &=& \phi(x) + \theta_{A}\phi_{A}(x+a_{A})
+\theta_{A}\theta_{B}\phi_{AB}(x+a_{A}+a_{B}) + \cdots, \\[2pt]
&=& \phi(x) + (-1)^{|\phi_{A}|}\phi_{A}(x-a_{A}) \theta_{A}
+\phi_{AB}(x-a_{A}-a_{B}) \theta_{A}\theta_{B} + \cdots,
\end{eqnarray}
where the component fields are chosen to be located in a symmetric manner.
The symbol $|\phi|$ again takes the value $+1(-1)$ for the bosonic (fermionic) 
component field $\phi$. 
All the superfield operators such as supercharges and supercovariant derivatives 
are accordingly represented in terms of the non-commutative $\theta_{A}$'s on 
the lattice.
The highlight of the formulation based on the relation (\ref{NC1}) 
is that the resulting lattice action can be manifestly invariant under all the SUSY
variations at least at tree level
if the lattice Leibniz rule conditions are satisfied \cite{DKKN1}.
On the other hand, it is rather non-trivial how to deal with 
the non-commutativity (\ref{NC1}) within the framework of quantum field theory.
One possible way to provide a field theoretically more rigid framework is 
to formulate the bosonic and fermionic coordinates entirely in terms of 
matrix representations and deal with 
the (super)fields as 
the functions of these (super)coordinate matrices.
A formulation presented in Ref. \cite{ADKS} may be placed
along this direction.

A somewhat alternative way to treat the non-commutativity (\ref{NC1}),
which is our starting point of this paper,
is to alter the notion of product and introduce a non-commutative product
in such a way to satisfy the relation,
\begin{eqnarray}
[x_{\mu},\theta_{A}]_{*} \ \equiv\ x_{\mu} * \theta_{A} - \theta_{A} * x_{\mu} 
&=& 2 (a_{A})_{\mu} \theta_{A}.
\label{NC2}
\end{eqnarray}
This type of non-commutativity treatment may be regarded as a certain analogue 
of the Moyal star product 
whose quantum implications have been extensively investigated 
({\it e.g.} \cite{Szabo} for a review). 
The principle we shall take 
is that all the non-commutativity should be originated 
from the nature of the product denoted as $*$ in (\ref{NC2}),
while the $x_{\mu}$ and $\theta_{A}$ should be treated as the ordinary
(discrete) bosonic coordinates and fermionic coordinates, respectively. 
In this framework, superfields can be expanded in terms of ordinary fermionic 
coordinates and ordinary functions,
\footnote{We shall take a convention to locate
all the component fields on a same site.
Note that superfields can 
be re-expressed in terms of the star product.
Using (\ref{NCfunc}) for instance, one has
$\Phi(x,\theta_{A}) = \phi(x) + \theta_{A}*\phi_{A}(x+a_{A})
+\theta_{A}*\theta_{B}*\phi_{AB}(x+a_{A}+a_{B}) + \cdots
= \phi(x) + (-1)^{|\phi_{A}|}\phi_{A}(x-a_{A}) * \theta_{A}
+\phi_{AB}(x-a_{A}-a_{B})* \theta_{A}*\theta_{B} + \cdots. $}
\begin{eqnarray}
\Phi(x,\theta_{A}) &=& \phi(x) + \theta_{A}\phi_{A}(x)
+\theta_{A}\theta_{B}\phi_{AB}(x) + \cdots, 
\end{eqnarray}
and thus may be served as path integration variables
such as in the standard SUSY treatment in the continuum spacetime 
\cite{Fujikawa-Lang,Ferrara-Piguet,Wess-Bagger,Piguet-Sibold}.
A star product of superfields, which is intrinsically non-commutative,
is defined in terms of $*$ subject to the relation (\ref{NC2}),
\begin{eqnarray}
\Phi_{1}(x,\theta_{A})*\Phi_{2}(x,\theta_{B})
\neq \Phi_{2}(x,\theta_{B})*\Phi_{1}(x,\theta_{A}).
\end{eqnarray}
An explicit representation of the star product satisfying the relation 
(\ref{NC2}) may be given by
\begin{eqnarray}
\Phi_{1}(x,\theta_{A})*\Phi_{2}(x,\theta_{B}) \equiv
\mu ( \mathcal{F}^{-1}_{L} \mathcal{F}_{R} \ 
\Phi_{1}(x,\theta_{A})\otimes \Phi_{2}(x,\theta_{B})),
\label{star_definition}
\end{eqnarray} 
with the elements $\mathcal{F}_{L}$ and $\mathcal{F}_{R}$, 
\begin{eqnarray}
\mathcal{F}_{L} &=& \exp \{{\sum_{\rho} \sum_{A} (a_{A})_{\rho}\theta_{A}
\frac{\partial}{\partial\theta_{A}} \otimes \partial_{\rho}}\}, 
\hspace{20pt}
\mathcal{F}_{R} \ =\ \exp\{{\sum_{\rho} \partial_{\rho} \otimes 
\sum_{A} (a_{A})_{\rho}\theta_{A}
\frac{\partial}{\partial\theta_{A}}}\}, \qquad
\label{F_definition}
\end{eqnarray}
where in (\ref{star_definition}) the symbol $\mu$ denotes the multiplication map,
$\mu(f\otimes g) \equiv fg$.
The operator $\theta_{A}\frac{\partial}{\partial\theta_{A}}$
in the r.h.s.'s of (\ref{F_definition})
should be understood as an adjoint operation,
\begin{eqnarray}
\theta_{A}\frac{\partial}{\partial\theta_{A}}
adj\ \theta_{A}\frac{\partial}{\partial\theta_{A}}
= [\theta_{A}\frac{\partial}{\partial\theta_{A}}, \ ],
\end{eqnarray}
together with the anti-commutation relation
between $\frac{\partial}{\partial\theta_{A}}$ and $\theta_{B}$,
\begin{eqnarray}
\{\frac{\partial}{\partial\theta_{A}},\theta_{B}\}= \delta_{AB}.
\label{anti-comm}
\end{eqnarray}
One may easily verify that the elements
$\mathcal{F}^{-1}_{L}$ and $\mathcal{F}_{R}$
independently give rise to the following relations,
\begin{eqnarray}
\mu(\mathcal{F}^{-1}_{L}\ \theta_{A} \otimes f(x)) &=& 
\theta_{A}f(x-a_{A}), \hspace{20pt}
\mu(\mathcal{F}^{-1}_{L}\ f(x) \otimes \theta_{A}) \ =\ 
f(x)\theta_{A}, \\[2pt]
\mu(\mathcal{F}_{R}\ \theta_{A} \otimes f(x)) &=& 
\theta_{A}f(x), \hspace{54pt}
\mu(\mathcal{F}_{R}\ f(x) \otimes \theta_{A}) \ =\ 
f(x+a_{A})\theta_{A}.
\end{eqnarray}
Accordingly, the star product defined in (\ref{star_definition})
would give,
\begin{eqnarray}
\theta_{A}*f(x) &=& \theta_{A}f(x-a_{A}), \hspace{20pt}
f(x)*\theta_{A} \ =\ f(x+a_{A})\theta_{A}.
\label{NCfunc}
\end{eqnarray} 
It is now obvious that the relation (\ref{NC2}) is satisfied 
just by taking $f(x)=x$.
It is important to notice that the star product (\ref{star_definition}) 
satisfies the associativity,
\begin{eqnarray}
\bigl( 
\Phi_{1}(x,\theta_{A})*\Phi_{2}(x,\theta_{B})
\bigr)
*\Phi_{3}(x,\theta_{C})
&=&
\Phi_{1}(x,\theta_{A})*
\bigl(
\Phi_{2}(x,\theta_{B})*\Phi_{3}(x,\theta_{C})
\bigr).
\label{associativity}
\end{eqnarray}
Also note that the star product of the super derivative operators 
$\frac{\partial}{\partial\theta_{A}}$
gives rise to the opposite shifts compared to (\ref{NCfunc}),
\begin{eqnarray}
\frac{\partial}{\partial\theta_{A}}*f(x) 
&=& \frac{\partial}{\partial\theta_{A}}f(x+a_{A}), \hspace{20pt}
f(x)*\frac{\partial}{\partial\theta_{A}} 
\ =\ f(x-a_{A})\frac{\partial}{\partial\theta_{A}},
\end{eqnarray}
with which one can verify that the anti-commutation relation (\ref{anti-comm}) 
holds also in the star product language, 
\begin{eqnarray}
\{\frac{\partial}{\partial\theta_{A}},\theta_{B}\}*f(x)
&\equiv& \frac{\partial}{\partial\theta_{A}}*\theta_{B}*f(x)
+\theta_{B}*\frac{\partial}{\partial\theta_{A}}*f(x) 
\ =\ \delta_{AB}f(x).
\end{eqnarray}
Although the combination of the elements
$(\mathcal{F}^{-1}_{L})^{t} (\mathcal{F}_{R})^{2-t}$
with an arbitrary parameter $t$ would always 
induce a non-commutative product which
satisfies the relation (\ref{NC2}),
in this paper we shall only consider the symmetric definition 
given in (\ref{star_definition}).

Having introduced a non-commutative product between the bosonic and 
fermionic coordinate of the superspace, $x_{\mu}$ and $\theta_{A}$,
we also require the same amount of 
non-commutativity between $x_{\mu}$ and Grassmann parameters $\xi_{A}$,
\begin{eqnarray}
\xi_{A}*f(x) &=& \xi_{A}f(x-a_{A}), \hspace{20pt}
f(x)*\xi_{A} \ =\ f(x+a_{A})\xi_{A},
\end{eqnarray}
such that the operation of $\xi_{A}\frac{\partial}{\partial\theta_{A}}$
would properly induce a derivative w.r.t. $\theta_{A}$
also in the language of the star product,
\begin{eqnarray}
\biggl[
\xi_{A}\frac{\partial}{\partial\theta_{A}},
\phi(x) + \theta_{B} \phi_{B}(x) + \cdots
\bigg]_{*}
&\equiv&
\xi_{A}\frac{\partial}{\partial\theta_{A}}
*\bigl( \phi(x) + \theta_{B} \phi_{B}(x) + \cdots \bigr) \nonumber \\
&&-\bigl( \phi(x) + \theta_{B} \phi_{B}(x) + \cdots \bigr) 
*\xi_{A}\frac{\partial}{\partial\theta_{A}} \nonumber \\[2pt]
&=& \delta_{AB}\xi_{A} \phi_{B}(x) +\cdots .
\end{eqnarray} 
This extension can easily be done by just replacing
the $\theta_{A}\frac{\partial}{\partial\theta_{A}}$
as $\xi_{A}\frac{\partial}{\partial\xi_{A}}+
\theta_{A}\frac{\partial}{\partial\theta_{A}}$ 
in the expressions (\ref{F_definition}).

Based on the above setup, we now proceed to define 
a lattice counterpart of SUSY algebra.
Since in the continuum spacetime, SUSY algebra 
can be regarded as 
a fermionic decomposition of differential operators,
one may naturally expect that a lattice counterpart of SUSY algebra
can be formally expressed as a fermionic decomposition of difference operators,
\begin{eqnarray}
\{Q_{A},Q_{B}\} &=& f^{\mu}_{AB}
d_{\mu},
\label{algebra_generic}
\end{eqnarray}
where $f^{\mu}_{AB}$ denotes a numerical coefficient, while 
$d_{\mu}$ 
denotes a ``formal" difference operator to be specified in the following. 
A SUSY transformation on the lattice 
is defined by a $*$ commutator between a supercharge $Q_{A}$
associated with a corresponding Grassmann parameter $\xi_{A}$
and a superfield $\Phi(x,\theta_{C})$,
\begin{eqnarray}
\delta_{A}\Phi(x,\theta_{C}) &\equiv& [ \xi_{A}Q_{A}, \Phi(x,\theta_{C}) ]_{*} 
\nonumber \\[2pt]
&=& \xi_{A}Q_{A} * \Phi(x,\theta_{C}) - \Phi(x,\theta_{C}) * \xi_{A}Q_{A}
\hspace{30pt}(A: \mathrm{no\ sum}). 
\label{SUSYvariation}
\end{eqnarray} 
A commutator of $\delta_{A}$ and $\delta_{B}$ would accordingly give,
\begin{eqnarray}
[\delta_{A},\delta_{B}]\Phi(x,\theta_{C}) &\equiv& 
[\xi_{A}Q_{A},[\xi_{B}Q_{B},\Phi(x,\theta_{C})]_{*}]_{*}
-[\xi_{B}Q_{B},[\xi_{A}Q_{A},\Phi(x,\theta_{C})]_{*}]_{*} \nonumber \\[2pt]
&=& -[\xi_{A}\xi_{B}\{Q_{A},Q_{B}\},\Phi(x,\theta_{C})]_{*} \nonumber \\[2pt]
&=& -\xi_{A}\xi_{B} f^{\mu}_{AB}
\bigl[
d_{\mu}\Phi(x-a_{A}-a_{B},\theta_{C})
-\Phi(x+a_{A}+a_{B},\theta_{C})d_{\mu}
\bigr],  \qquad
\label{commutator}
\end{eqnarray}
where in the last line we substitute the relation (\ref{algebra_generic}).
One can notice here that the SUSY variation defined in terms of the 
star product (\ref{SUSYvariation}) 
automatically induces a difference operation in the last line, provided
either of the following conditions holds,
\begin{eqnarray}
a_{A}+a_{B} &=& +n_{\mu} \hspace{20pt} \mathrm{and} \hspace{20pt}
d_{\mu} = -\frac{1}{2},
\label{Leibniz_cond1}
\\[2pt]
a_{A}+a_{B} &=& -n_{\mu} \hspace{20pt} \mathrm{and} \hspace{20pt}
d_{\mu} = +\frac{1}{2},
\label{Leibniz_cond2}
\end{eqnarray}
where $n_{\mu}$ is denoting a unit vector whose component is defined by
$(n_{\mu})_{\nu}=\delta_{\mu\nu}$.
One may verify that 
the last line of the expression (\ref{commutator}) 
has the desired continuum limit
if the condition either (\ref{Leibniz_cond1}) or (\ref{Leibniz_cond2})
is satisfied, 
\begin{eqnarray}
-\xi_{A}\xi_{B} f^{\mu}_{AB}
\frac{1}{2}\bigl[
\Phi(x+n_{\mu},\theta_{C})
-\Phi(x-n_{\mu},\theta_{C})
\bigr]
&\rightarrow&
-\xi_{A}\xi_{B} f^{\mu}_{AB} 
\partial_{\mu} \Phi(x,\theta_{C}).
\end{eqnarray} 
The superspace expressions for the supercharges satisfying (\ref{algebra_generic})
are readily given by, for instance,
\begin{eqnarray}
Q_{A} &=& \frac{\partial}{\partial \theta_{A}}
+\frac{1}{2}f^{\mu}_{AB}\theta_{B}d_{\mu}, \hspace{30pt}
Q_{B} \ =\ \frac{\partial}{\partial \theta_{B}}
+\frac{1}{2}f^{\mu}_{AB}\theta_{A}d_{\mu}.
\end{eqnarray}
Remark here that the difference operation in this framework
is generated  by the non-commutative nature of the star product
and thus the lattice SUSY algebra given in (\ref{algebra_generic})
is actually describing a ``reduced" algebra in the sense that
the differential operator $\partial_{\mu}$ in the continuum SUSY algebra
is replaced by a constant value 
$d_{\mu} \ =\ \pm\frac{1}{2}$
in its lattice counterpart.
The conditions (\ref{Leibniz_cond1}) and 
(\ref{Leibniz_cond2}) coincide with the Leibniz rule conditions
introduced in the previous works \cite{DKKN1,DKKN2,DKKN3} where it was shown
that Dirac-K\"ahler twisted SUSY algebra
of $\cN=D=2$, $\cN=4\ D=3$ and $\cN=D=4$ can satisfy these conditions.
Although the above star product formulation is formally available in 
any of these cases,
in this paper we mainly concentrate on the simplest but non-trivial 
case, $\cN=D=2$.
It is also important to remark that
a SUSY variation for a star product of superfields 
$\Phi_{1}*\Phi_{2}$
satisfies the following Leibniz rule
thanks to the associativity (e.g. (\ref{associativity})),
\begin{eqnarray}
\delta_{A}(\Phi_{1} * \Phi_{2})
&=& [\xi_{A}Q_{A},\Phi_{1}*\Phi_{2}]_{*} \nonumber \\[2pt]
&=& \xi_{A}Q_{A} * \Phi_{1} * \Phi_{2} 
-  \Phi_{1} * \Phi_{2} * \xi_{A}Q_{A} \nonumber \\[2pt]
&=& (\delta_{A}\Phi_{1} + \Phi_{1} * \xi_{A}Q_{A} ) * \Phi_{2}
- \Phi_{1} * (\xi_{A}Q_{A}*\Phi_{2} - \delta_{A}\Phi_{2}) \nonumber \\[2pt]
&=& \delta_{A}\Phi_{1}* \Phi_{2} + \Phi_{1} * \delta_{A} \Phi_{2},  
\label{Q_associativity}
\end{eqnarray}
while a SUSY variation for an ordinary product 
$\Phi_{1}\Phi_{2}$
does not,
\begin{eqnarray}
\delta_{A}(\Phi_{1}  \Phi_{2})
\neq
\delta_{A}\Phi_{1} \Phi_{2} + \Phi_{1}  \delta_{A} \Phi_{2},  
\end{eqnarray} 
since in general we have,
$(A*B)C \neq A*(BC)$.
From these properties, one can immediately see that
any lattice action which is to be invariant under 
the SUSY variation (\ref{SUSYvariation})
should be expressed in terms of the star product.

\section{$\cN=D=2$ Dirac-K\"ahler Twisted SUSY Algebra 
\& Wess-Zumino Model}

\indent

In this section, as a simple and explicit example,
we introduce the $\cN=D=2$ twisted SUSY algebra
and see how the lattice counterpart of 
the algebra can be realized 
in accordance with the lattice Leibniz rule conditions.
We introduce a chiral and anti-chiral superfields 
to construct a $\cN=D=2$ twisted Wess-Zumino model on a lattice.
The superfield propagators and the three-point
vertex functions shall be explicitly derived 
by means of the lattice superfield method.


\subsection{$\cN=D=2$ twisted SUSY algebra on a lattice}

\indent

Let us first introduce the $\cN=D=2$ SUSY algebra in the continuum spacetime
which is given by
\begin{eqnarray}
&&\{Q_{\alpha i},\overline{Q}_{j\beta}\} 
\ =\ 2i\delta_{ij}(\gamma_{\mu})_{\alpha\beta}\partial_{\mu}, 
\label{N=D=2_algebra}
\\[2pt]
[J,Q_{\alpha i}] &=& \frac{i}{2}(\gamma_{5})_{\alpha\beta}Q_{\beta i}, 
\hspace{20pt}
[R,Q_{\alpha i}] \ =\ \frac{i}{2}(\gamma_{5})_{ij}Q_{\alpha j},  \\[2pt]
[J,\partial_{\mu}] &=& i\epsilon_{\mu\nu}\partial_{\nu}, 
\hspace{20pt}
[R,\partial_{\mu}] \ =\ [\partial_{\mu}, Q_{\alpha i}] 
\ =\ [\partial_{\mu}, \partial_{\nu}] \ =
\ [J,R] \ =\ 0. 
\end{eqnarray}
where the indices $\alpha, \beta=1\ \mathrm{or}\ 2$ 
and $i,j =1\ \mathrm{or}\ 2$ denote spin and $\cN=2$ internal indices, respectively. 
The gamma matrices in two dimensions can be represented by Pauli matrices,
\begin{eqnarray}
\gamma_{1}=\sigma_{3}, \ \ \ \gamma_{2}=\sigma_{1}, \ \ \  
\gamma_{5}=\gamma_{1}\gamma_{2} = i\sigma_{2}.
\label{Pauli_matrices}
\end{eqnarray}
The conjugate supercharge 
$\overline{Q} \equiv Q^{\dagger}$ 
can be taken as Majorana,
subject to 
the condition
$\overline{Q}_{i\alpha} = (Q^{T})_{i\alpha} = Q_{\alpha i}$.
$J$ and $R$ denotes the generators for the $SO(2)$ Lorentz 
and $SO(2)$ internal rotations, respectively.
The twisted form of the algebra (\ref{N=D=2_algebra})
is given in terms of the Dirac-K\"ahler expansion of 
the supercharge,
\begin{eqnarray}
Q_{\alpha i} &=& ({\bf 1}Q + \gamma_{\mu}Q_{\mu} + \gamma_{5}\tilde{Q})_{\alpha i},
\end{eqnarray}
as
\begin{eqnarray}
\{Q,Q_{\mu}\} &=& i\partial_{\mu}, \hspace{30pt}
\{\tilde{Q},Q_{\mu}\} \ =\ -i\epsilon_{\mu\nu}\partial_{\nu}, 
\hspace{30pt}
\{others\} \ =\ 0.
\label{N=D=2_twisted_algebra}
\end{eqnarray}
The twisted supercharges $(Q,Q_{\mu},\tilde{Q})$ transform
as a scalar, a vector and a (pseudo-)scalar under 
the twisted Lorentz generator $J'\equiv J+R$, respectively,
in the continuum spacetime,
\begin{eqnarray}
[J',Q] &=& 0,
\hspace{30pt}
[J',Q_{\mu}] \ =\ i\epsilon_{\mu\nu}Q_{\nu},
\hspace{30pt}
[J',\tilde{Q}] \ =\ 0.
\end{eqnarray}

\begin{figure}
\begin{center}
\includegraphics[width=40mm]{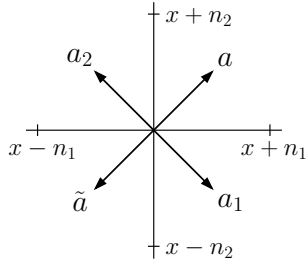}
\caption{Symmetric choice of $a_{A}$}
\label{symm_a}
\end{center}
\end{figure}

It was first pointed out in Ref. \cite{DKKN1} that
the algebra which can be realized on the lattice is not the form
of (\ref{N=D=2_algebra}) but the form of (\ref{N=D=2_twisted_algebra})
since only the Dirac-K\"ahler twisted form of SUSY
algebra has a chance to
satisfy the Leibniz rule condition.
In the present context of the formulation, the lattice counterpart 
of the algebra (\ref{N=D=2_twisted_algebra})
can be expressed as,
\begin{eqnarray}
\{Q,Q_{\mu}\} &=& id^{+}_{\mu},
\hspace{30pt}
\{\tilde{Q},Q_{\mu}\} \ =\ -i\epsilon_{\mu\nu}d^{-}_{\nu}, 
\hspace{30pt}
\{others\} \ =\ 0,
\label{N=D=2_lattice_algebra}
\end{eqnarray}
with the corresponding Leibniz rule conditions,
\begin{eqnarray}
a+a_{\mu} &=& +n_{\mu}, \hspace{20pt}
\tilde{a}+a_{\mu} \ =\ -|\epsilon_{\mu\nu}|n_{\nu}.
\label{N=D=2_Leibniz_cond}
\end{eqnarray}
The ``formal" difference operators $d_{\mu}^{+}$ and $d_{\mu}^{-}$
in the r.h.s. of (\ref{N=D=2_twisted_algebra}) 
take the value of $-\frac{1}{2}$ and $+\frac{1}{2}$, respectively,
as is required by the conditions (\ref{Leibniz_cond1}) and (\ref{Leibniz_cond2}).
The conditions (\ref{N=D=2_Leibniz_cond}) 
are satisfied by the following generic solutions 
with the one vector arbitrariness,
\begin{eqnarray}
a &=& (arbitrary), \hspace{20pt}
a_{\mu} \ =\ +n_{\mu} -a, \hspace{20pt}
\tilde{a} \ =\ -n_{1} - n_{2} +a.
\end{eqnarray}
Note that the sum of all the shift parameters vanishes
regardless of the one vector arbitrariness,
\begin{eqnarray}
a + \ta + a_{1} + a_{2} &=& 0.
\label{a_vanishing_sum}
\end{eqnarray}  
In this paper, 
mainly because of the advantages in dealing with the superfield formulation, 
we shall take  the symmetric choice 
which is depicted in Fig. \ref{symm_a},
\begin{eqnarray}
a &=& (+\frac{1}{2},+\frac{1}{2}), \hspace{10pt}
a_{1} \ =\  (+\frac{1}{2},-\frac{1}{2}), \hspace{10pt}
a_{2} \ =\ (-\frac{1}{2},+\frac{1}{2}), \hspace{10pt}
\tilde{a} \ =\ (-\frac{1}{2},-\frac{1}{2}). 
\label{a_symmetric_choice}
\end{eqnarray}
Superspace expressions for the supercharges 
$Q_{A}=(Q,Q_{\mu},\tilde{Q})$ which satisfy 
(\ref{N=D=2_lattice_algebra}) are given by,
\begin{eqnarray}
Q &=& \frac{\partial}{\partial \theta} 
+ \frac{i}{2}\theta_{\mu}d^{+}_{\mu}, \label{Q1} 
\hspace{30pt}
Q_{\mu} \ =\ \frac{\partial}{\partial \theta_{\mu}}
+ \frac{i}{2}\theta d^{+}_{\mu}
- \frac{i}{2}\tilde{\theta}\epsilon_{\mu\nu}d^{-}_{\nu}, \\[2pt]
\tilde{Q} &=& \frac{\partial}{\partial\tilde{\theta}} 
-\frac{i}{2}\theta_{\mu} \epsilon_{\mu\nu} d^{-}_{\nu}.
\label{Q3}
\end{eqnarray}
We also introduce the following supercovariant derivatives 
$D_{A}=(D,D_{\mu},\tilde{D})$, 
\begin{eqnarray}
D &=& \frac{\partial}{\partial \theta} 
- \frac{i}{2}\theta_{\mu}d^{+}_{\mu}, \label{D1} 
\hspace{30pt}
D_{\mu} \ =\ \frac{\partial}{\partial \theta_{\mu}}
- \frac{i}{2}\theta d^{+}_{\mu}
+ \frac{i}{2}\tilde{\theta}\epsilon_{\mu\nu}d^{-}_{\nu}, \label{D2} \\[2pt]
\tilde{D} &=& \frac{\partial}{\partial\tilde{\theta}} 
+\frac{i}{2}\theta_{\mu} \epsilon_{\mu\nu} d^{-}_{\nu}, \label{D3}
\end{eqnarray}
which anti-commute with the supercharges,
$\{Q_{A},D_{B}\}=0$, and satisfy the algebra,
\begin{eqnarray}
\{D,D_{\mu}\} &=& -id^{+}_{\mu},
\hspace{30pt}
\{\tilde{D},D_{\mu}\} \ =\ i\epsilon_{\mu\nu}d^{-}_{\nu}, 
\hspace{30pt}
\{others\} \ =\ 0.
\label{N=D=2_lattice_D_algebra}
\end{eqnarray}

A chiral superfield $\Phi(x,\theta_{A},\otheta_{A})$ and
an anti-chiral superfield $\overline{\Phi}(x,\theta_{A},\otheta_{A})$
on the lattice
can be defined in terms of the supercovariant derivatives (\ref{D1})-(\ref{D3}),
\begin{eqnarray}
[\xi D, \Phi(x,\theta_{A},\otheta_{A})]_{*} 
&=& [\tilde{\xi} \tilde{D}, \Phi(x,\theta_{A},\otheta_{A})]_{*} 
\ =\ 0, \label{chiral_cond_lat1}\\[2pt]
[\xi_{\mu}D_{\mu}, \overline{\Phi}(x,\theta_{A},\otheta_{A})]_{*} &=& 0 \hspace{20pt}
\mathrm{(\mu \ :\ no\ sum)}, \label{chiral_cond_lat2}
\end{eqnarray}
here and in the following 
we shall represent
the arguments of the superfields by
$\theta_{A}\equiv\theta_{\mu}=(\theta_{1},\theta_{2})$ 
and $\otheta_{A}\equiv (\theta,\ttheta)$.
In solving these conditions, 
it is convenient to introduce 
the following difference expansion operator,
\begin{eqnarray}
U \equiv e^{-\frac{i}{2}(\theta\theta_{\mu}d^{+}_{\mu}
-\epsilon_{\mu\nu}\tilde{\theta}\theta_{\mu}d^{-}_{\nu})}_{*},
\label{U_def}
\end{eqnarray}
where the exponential with the star is defined by
\begin{eqnarray}
e^{X}_{*} &\equiv& 1+ X + \frac{1}{2}X*X + \frac{1}{3!}X*X*X + \cdots.
\end{eqnarray}
The supercovariant derivatives (\ref{D1}-\ref{D3}) can be expressed
in terms of $U$ as,
\begin{eqnarray}
\xi D &=& U^{-1} * \xi\frac{\partial}{\partial\theta} * U, \hspace{20pt}
\tilde{\xi} \tilde{D} \ =\ 
U^{-1} * \tilde{\xi}\frac{\partial}{\partial\tilde{\theta}} * U, \\[2pt]
\xi_{\mu}D_{\mu} & = & U * \xi_{\mu}\frac{\partial}{\partial\theta_{\mu}} * U^{-1} 
\hspace{20pt} (\mathrm{\mu :\ no\ sum}),
\end{eqnarray}
with which one may rewrite the chiral and anti-chiral conditions 
(\ref{chiral_cond_lat1})-(\ref{chiral_cond_lat2}) as
\begin{eqnarray}
[\xi\frac{\partial}{\partial\theta},\Phi'(x,\theta_{A},\otheta_{A})]_{*} 
&=& [\tilde{\xi}\frac{\partial}{\partial\tilde{\theta}},
\Phi'(x,\theta_{A},\otheta_{A})]_{*}
\ =\ 0, 
\label{Chiral_cond_modified1}
\\[2pt]
[\xi_{\mu}\frac{\partial}{\partial\theta_{\mu}},
\overline{\Phi}'(x,\theta_{A},\otheta_{A})]_{*}
&=& 0 \hspace{20pt} \mathrm{(\mu \ :\ no\ sum)},
\label{Chiral_cond_modified2}
\end{eqnarray}
where $\Phi'(x,\theta_{A},\otheta_{A})$ 
and $\overline{\Phi}'(x,\theta_{A},\otheta_{A})$ are defined by
\begin{eqnarray}
\Phi'(x,\theta_{A},\otheta_{A}) 
&=& U * \Phi(x,\theta_{A},\otheta_{A}) * U^{-1},  \hspace{10pt}
\overline{\Phi}'(x,\theta_{A},\otheta_{A}) 
\ = \ U^{-1}* \overline{\Phi}(x,\theta_{A},\otheta_{A}) * U. \qquad
\label{phi_phi_prime}
\end{eqnarray}
One can easily solve the conditions
(\ref{Chiral_cond_modified1}) and (\ref{Chiral_cond_modified2})
and get the following expressions for
$\Phi'$ and $\oPhi'$,
\begin{eqnarray}
\Phi'(x,\theta_{A},\otheta_{A}) &=& \Phi'(x,\theta_{A}) \ =\
\phi(x) + \theta_{\mu}\psi_{\mu}(x) + \theta_{1}\theta_{2}\tilde{\phi}(x), 
\label{phi_prime1}   \\[2pt]
\overline{\Phi}'(x,\theta_{A},\otheta_{A}) 
&=& \overline{\Phi}'(x,\otheta_{A}) \ =\
\varphi(x) + \theta\chi(x) + \tilde{\theta}\tilde{\chi} (x)
+ \theta\tilde{\theta}\tilde{\varphi}(x),
\label{phi_prime2}
\end{eqnarray}
where $(\phi,\tilde{\phi},\varphi,\tilde{\varphi})$
and $(\psi_{1},\psi_{2},\chi,\tchi)$
are denoting the bosonic and  fermionic 
component fields, respectively.
Then the expressions of the chiral and anti-chiral superfields 
$(\Phi,\oPhi)$ can be found as, 
\begin{eqnarray}
\Phi(x,\theta_{A},\otheta_{A}) &=& U^{-1}* \Phi'(x,\theta_{A}) * U \nonumber \\[2pt]
&=& \phi(x) + \theta_{\mu}\psi_{\mu}(x) + \theta_{1}\theta_{2}\tilde{\phi}(x) 
+ \frac{i}{2}\theta\theta_{\mu}\Delta_{\mu}\phi(x)
-\frac{i}{2}\epsilon_{\mu\nu}\tilde{\theta}\theta_{\mu}\Delta_{\nu}\phi(x) 
\nonumber \\[2pt]
&&+\frac{i}{2}\epsilon_{\mu\nu}\theta\theta_{1}\theta_{2}\Delta_{\mu}\psi_{\nu}(x)
-\frac{i}{2}\tilde{\theta}\theta_{1}\theta_{2}\Delta_{\mu}\psi_{\mu}(x) 
+\frac{1}{4}\theta\tilde{\theta}\theta_{1}\theta_{2}
\Delta_{\mu}\Delta_{\mu}\phi(x), \\[6pt]
\oPhi(x,\theta_{A},\otheta_{A}) 
&=& U * \oPhi'(x,\otheta_{A}) * U^{-1} \nonumber \\[2pt]
&=& 
\varphi(x) + \theta\chi(x) + \tilde{\theta}\tilde{\chi} (x)
+ \theta\tilde{\theta}\tilde{\varphi}(x)
-\frac{i}{2}\theta\theta_{\mu}\Delta_{\mu}\varphi(x)
+\frac{i}{2}\epsilon_{\mu\nu}\tilde{\theta}\theta_{\mu}\Delta_{\nu}\varphi(x) 
\nonumber \\[2pt]
&&+\frac{i}{2}\epsilon_{\mu\nu}\theta\tilde{\theta}\theta_{\mu}\Delta_{\nu}\chi(x)
+\frac{i}{2}\theta\tilde{\theta}\theta_{\mu}\Delta_{\mu}\tilde{\chi}(x)
+\frac{1}{4}\theta\tilde{\theta}\theta_{1}\theta_{2}
\Delta_{\mu}\Delta_{\mu}\varphi(x),
\end{eqnarray}
where $\Delta_{\mu}$ is denoting a symmetric difference operator,
$\Delta_{\mu}f(x)=\frac{1}{2}(f(x+n_{\mu})-f(x-n_{\mu}))$.
In deriving these expressions we made use of 
a star product version of the exponential formula,
\begin{eqnarray}
e^{X}_{*}*Y*e^{-X}_{*} &=& Y +[X,Y]_{*} +\frac{1}{2!}[X,[X,Y]_{*}]_{*} + \cdots,
\end{eqnarray} 
which obeys from the associativity of the star product.

The $\cN=D=2$ twisted SUSY transformation laws for the component fields 
follow from the definition (\ref{SUSYvariation}).
As for the chiral superfield $\Phi(x,\theta_{A},\otheta_{A})$, we have
\begin{eqnarray}
\delta_{\xi_{A}} \Phi(x,\theta_{B},\otheta_{B}) 
&=& U^{-1} * \delta_{\xi_{A}} \Phi'(x,\theta_{B}) * U \nonumber \\
&=& [\xi_{A}Q_{A}, U^{-1} * \Phi'(x,\theta_{B}) * U]_{*},
\end{eqnarray}
from which it obeys
\begin{eqnarray}
\delta_{\xi_{A}} \Phi'(x,\theta_{B})
&=& [\xi_{A}Q'_{A}, \Phi'(x,\theta_{B})]_{*},
\end{eqnarray}
where $\xi_{A}Q'_{A} \equiv U*\xi_{A}Q_{A}*U^{-1}$ are given by
\begin{eqnarray}
\xi Q' &=& \xi  \frac{\partial}{\partial\theta}
+i \xi \theta_{\mu}d^{+}_{\mu}, \hspace{20pt}
\xi_{\mu}Q'_{\mu} \ =\  \xi_{\mu}\frac{\partial}{\partial\theta_{\mu}} 
\ \ \ \ (\mu:\mathrm{no\ sum}),  \\
\tilde{\xi} \tilde{Q}' & =&  \tilde{\xi}  \frac{\partial}{\partial\tilde{\theta}}
-i \tilde{\xi} \theta_{\mu}\epsilon_{\mu\nu}d^{-}_{\nu}.
\end{eqnarray}
In a similar way for the anti-chiral superfield $\oPhi(x,\theta_{A},\otheta_{A})$,
we have
\begin{eqnarray}
\delta_{\xi_{A}} \oPhi'(x,\otheta_{A})
&=& [\xi_{A}Q''_{A}, \oPhi'(x,\otheta_{A})]_{*},
\end{eqnarray}
where $\xi_{A}Q''_{A} \equiv U^{-1}*\xi_{A}Q_{A}*U$ are given by
\begin{eqnarray}
\xi Q'' &=& \xi\frac{\partial}{\partial\theta}, \hspace{20pt} 
\xi_{\mu}Q''_{\mu} \ =\
\xi_{\mu} \frac{\partial}{\partial\theta_{\mu}}
+ i\xi_{\mu}\theta d^{+}_{\mu}
- i\xi_{\mu}\tilde{\theta} \epsilon_{\mu\nu} d^{-}_{\nu}
\ \ \ \ (\mu:\mathrm{no\ sum}), \\
\tilde{\xi} \tilde{Q}'' & =& \tilde{\xi}\frac{\partial}{\partial\tilde{\theta}}. 
\end{eqnarray}
The $\cN=D=2$ twisted SUSY transformation laws for the component fields
are summarized in 
Table \ref{SUSYtrans_laws}.
One can verify that the component-wise SUSY variations 
form the following off-shell closed algebra,
\begin{eqnarray}
[\delta_{\xi},\delta_{\xi_{\rho}}] \varphi_{A} 
&=& + i\xi\xi_{\rho}\Delta_{\rho} \varphi_{A} \hspace{20pt}
(\mathrm{\rho : no\ sum}), \\[2pt]
[\delta_{\tilde{\xi}},\delta_{\xi_{\rho}}] \varphi_{A}
&=& -i\tilde{\xi}\xi_{\rho}\epsilon_{\rho\sigma}\Delta_{\sigma} \varphi_{A}
\hspace{20pt}
(\mathrm{\rho : no\ sum}), \\[2pt]
[ others ] \varphi_{A} &=& 0, 
\end{eqnarray}
where $\varphi_{A}$ denotes any of the component field 
$(\phi,\psi_{\mu},\tilde{\phi},\varphi,\chi,\tilde{\chi},\tilde{\varphi})$.

\begin{table}
\begin{center}
\renewcommand{\arraystretch}{1.4}
\renewcommand{\tabcolsep}{10pt}
\begin{tabular}{c|ccc}
\hline
& $\delta_{\xi}$ & $\delta_{\xi_{\rho}}$ & $\delta_{\tilde{\xi}}$ \\ \hline
$\phi(x)$ & $0$  & $\xi_{\rho}\psi_{\rho}(x)$ & $0$ \\
$\psi_{\mu}(x)$ & $-i\xi\Delta_{\mu}\phi(x)$  
& $\epsilon_{\mu\rho}\xi_{\rho}\tilde{\phi}(x)$  
& $i\epsilon_{\mu\rho}\tilde{\xi}\Delta_{\rho}\phi(x)$ \\
$\tilde{\phi}(x)$ & $i\xi\epsilon_{\rho\sigma}\Delta_{\rho}\psi_{\sigma}(x)$ 
& $0$ & $-i\tilde{\xi}\Delta_{\rho}\psi_{\rho}(x)$ \\ \hline
$\varphi(x)$ & $\xi\chi(x)$ & $0$  
& $\tilde{\xi}\tilde{\chi}(x)$ \\
$\chi(x)$ & $0$ & $-i\xi_{\rho}\Delta_{\rho}\varphi(x)$ 
& $\tilde{\xi}\tilde{\varphi}(x)$ \\
$\tilde{\chi}(x)$ & $-\xi\tilde{\varphi}(x)$ 
& $i\epsilon_{\rho\sigma}\xi_{\rho}\Delta_{\sigma}\varphi(x)$ & $0$ \\
$\tilde{\varphi}(x)$ & $0$  
& $i\xi_{\rho}\Delta_{\rho}\tilde{\chi}(x) 
+i\xi_{\rho}\epsilon_{\rho\sigma}\Delta_{\sigma}\chi(x)$ & $0$ \\ \hline  
\end{tabular}
\caption{$\cN=D=2$ twisted SUSY transformation laws 
on the lattice for the component fields
$(\phi,\psi_{\mu},\tilde{\phi})$ and $(\varphi,\chi,\tilde{\chi},\tilde{\varphi})$.}
\label{SUSYtrans_laws}
\end{center}
\end{table}


\subsection{Twisted SUSY invariant action at tree level}

\indent

In terms of the chiral and anti-chiral superfields $\Phi(x,\theta_{A})$
and $\oPhi(x,\theta_{A})$,
one can generally construct a $\cN=D=2$ twisted SUSY invariant action
on the lattice as 
\begin{eqnarray}
S &=& \sum_{x} \int d^{4}\theta \ K_{*}(\oPhi,\Phi)
+ \sum_{x} \int d^{2}\theta \ F_{*}(\Phi)
+ \sum_{x} \int d^{2}\overline{\theta} \ \overline{F}_{*}(\oPhi)
\end{eqnarray}
where the Grassmann measures are defined by
$d^{4}\theta \equiv d\theta d\tilde{\theta}d\theta_{1}d\theta_{2}$,
$d^{2}\theta \equiv d\theta_{2}d\theta_{1}$
and $d^{2}\overline{\theta} \equiv d\tilde{\theta}d\theta$,
and each Grassmann integration is defined by
$\int d\theta_{A}\ 1 = 0$,
$\int d\theta_{A}\theta_{B}=\delta_{AB}$. 
The symbol $K_{*}$ 
is denoting any functions of 
the chiral and anti-chiral superfields in terms of the star product,
while the symbols 
$F_{*}$ and $\overline{F}_{*}$ are respectively denoting any functions of 
the chiral superfields and anti-chiral superfields
again in terms of the star product.
Since we take the symmetric choice of $a_{A}$ 
(\ref{a_symmetric_choice}),  
the summation over $x$ should cover 
(int, int) sites as well as (half-int, half-int) sites,
\begin{eqnarray}
\sum_{x} &=& \sum_{(m_{1},m_{2})} + \sum_{(m_{1}+\frac{1}{2},m_{2}+\frac{1}{2})}
\hspace{20pt}
(m_{1},m_{2} : \mathrm{integers}).
\label{x_sum}
\end{eqnarray}
The necessity of summing over both (int,int) sites and (half-int,half-int) sites
stems from the fact that in the symmetric choice of $a_{A}$
the lattice SUSY transformation
can essentially be regarded as a mapping from 
(int,int) to (half-int,half-int) or vice versa, 
due to the non-commutativity associated with $\theta_{A}$ and $\xi_{A}$.
The simplest action with mass and interaction terms
can be given by a $\cN=D=2$ Dirac-K\"ahler twisted analog of the Wess-Zumino model 
\cite{Wess-Zumino},
\begin{eqnarray}
K_{*} = \oPhi * \Phi, \hspace{20pt}
F_{*} = \frac{m}{2} \Phi * \Phi + \frac{g}{3!} \Phi * \Phi * \Phi, \hspace{20pt}
\overline{F}_{*} 
= \frac{m}{2} \oPhi * \oPhi + \frac{g}{3!} \oPhi * \oPhi * \oPhi.
\end{eqnarray}
The total action $S_{WZ}$ is accordingly given by,
\begin{eqnarray}
S_{WZ} &=& S_{kin} + S_{mass} + S_{int},
\label{total_action}
\end{eqnarray}
where the kinetic terms $S_{kin}$, the mass terms $S_{mass}$ 
and the interaction terms 
$S_{int}$ are expressed in terms of the component fields as,
\begin{eqnarray}
S_{kin} &=& \sum_{x} \int d^{4}\theta \ \oPhi(x,\theta_{A},\otheta_{A}) 
* \Phi(x,\theta_{A},\otheta_{A}) 
\nonumber \\[2pt]
&=& \sum_{x} \biggl[
\varphi(x)\Delta_{\mu}\Delta_{\mu}\phi(x) 
+ i\chi(x)\Delta_{\mu}\psi_{\mu}(x)
+ i\epsilon_{\mu\nu}\tilde{\chi}(x)\Delta_{\mu}\psi_{\nu}(x)
+ \tilde{\varphi}(x)\tilde{\phi}(x)
\biggr], \qquad
\label{kinetic_terms}
\end{eqnarray}
\begin{eqnarray}
S_{mass} &=& \frac{m}{2} \sum_{x} \biggl[
\int d^{2}\theta \
\Phi(x,\theta_{A},\otheta_{A})*\Phi(x,\theta_{A},\otheta_{A}) 
+ \int d^{2}\overline{\theta} \
\oPhi(x,\theta_{A},\otheta_{A})*\oPhi(x,\theta_{A},\otheta_{A})  \biggr] \nonumber \\
&=& \frac{m}{2} \sum_{x} \biggl[
\phi(x)\tilde{\phi}(x) + \tilde{\phi}(x)\phi(x)
-\epsilon_{\mu\nu}\psi_{\mu}(x)\psi_{\nu}(x)  \nonumber \\ 
&& + \varphi(x)\tilde{\varphi}(x) + \tilde{\varphi}(x)\varphi(x)
-\chi(x)\tilde{\chi}(x) + \tilde{\chi}(x)\chi(x)
\biggr], 
\label{mass_terms}
\end{eqnarray}
\begin{eqnarray}
S_{int} &=&  \frac{g}{3!} \sum_{x} \biggl[
\int d^{2}\theta \
\Phi(x,\theta_{A},\otheta_{A}) * \Phi(x,\theta_{A},\otheta_{A}) 
* \Phi(x,\theta_{A},\otheta_{A}) \nonumber \\[2pt] 
&&+ \int d^{2}\overline{\theta} \
\oPhi(x,\theta_{A},\otheta_{A}) * \oPhi(x,\theta_{A},\otheta_{A}) 
* \oPhi(x,\theta_{A},\otheta_{A}) 
\biggr] \nonumber \\
&=& \frac{g}{3!} \sum_{x} \biggl[
-\epsilon_{\mu\nu}  \phi(x)\psi_{\mu}(x+a_{\nu})\psi_{\nu}(x-a_{\mu}) 
-\epsilon_{\mu\nu} \psi_{\mu}(x+a_{\nu})\psi_{\nu}(x-a_{\mu})\phi(x)  
 \nonumber \\
&& -\epsilon_{\mu\nu} \psi_{\mu}(x-a_{\nu})\phi(x)\psi_{\nu}(x+a_{\mu})
+ \phi(x)\phi(x)\tilde{\phi}(x)
+ \phi(x)\tilde{\phi}(x)\phi(x)
+ \tilde{\phi}(x)\phi(x)\phi(x) \nonumber \\[7pt]
&&- \varphi(x)\chi(x+\ta)\tchi(x-a)
+ \varphi(x)\tchi(x+a)\chi(x-\ta)
-\chi(x+\ta)\tchi(x-a)\varphi(x) \nonumber \\[7pt]
&&+\tchi(x+a)\chi(x-\ta)\varphi(x)
-\chi(x-\ta)\varphi(x)\tchi(x+a)
+\tchi(x-a)\varphi(x)\chi(x+\ta) \nonumber \\
&& + \varphi(x)\varphi(x)\tilde{\varphi}(x)
+ \varphi(x)\tilde{\varphi}(x)\varphi(x)
+ \tilde{\varphi}(x)\varphi(x)\varphi(x) 
\biggr],
\label{int_terms}
\end{eqnarray}
where we used the symmetric nature of the shift parameters 
(\ref{a_symmetric_choice}), $a_{1}+a_{2}=0$, $a+\ta=0$.
One should notice that the kinetic terms (\ref{kinetic_terms}) 
and mass terms (\ref{mass_terms})
consist of the component fields living only
on (int, int) sites or (half-int, half-int) sites,
which implies that the field propagations are restricted 
only from (int, int) to (int, int) or from (half-int, half-int) 
to (half-int, half-int) sites.
On the other hand, the interaction terms (\ref{int_terms})
involve both (int, int) and (half-int, half-int) sites.
In particular, the Yukawa coupling terms consist of 
the bosonic field at an (int, int) site 
and the fermionic fields at (half-int, half-int) sites
or vice versa.

The twisted $\cN=D=2$ SUSY invariant nature of the total action 
is manifest from their superfield expressions,
since the star products of the superfields
satisfy the Leibniz rule (\ref{Q_associativity})
which is also valid for a product of three superfields,
\begin{eqnarray}
\delta_{A} (\Phi_{1}*\Phi_{2}*\Phi_{3})
&=& \delta_{A}\Phi_{1} *\Phi_{2}*\Phi_{3}
+ \Phi_{1} * \delta_{A}\Phi_{2} * \Phi_{3}
+ \Phi_{1} * \Phi_{2} * \delta_{A}\Phi_{3}. 
\end{eqnarray} 
It is important to note that the above star product nature
is implicitly inherited also in the component-wise expressions.
Actually, in order to be consistent with the star product nature 
of the superfields which preserves the Leibniz rule,
one needs to recognize that 
the SUSY variation of each component field should be associated with 
the star product, for example,
\begin{eqnarray}
\delta_{A}(\phi(x)\phi(x)\tilde{\phi}(x))
&=& \delta_{A}\phi(x)*\phi(x)\tilde{\phi}(x)
+ \phi(x) * \delta_{A}\phi(x) * \tilde{\phi}(x)
+ \phi(x) \phi(x) * \delta_{A}\tilde{\phi}(x). \nonumber \\
\end{eqnarray}
Since the component-wise SUSY variations are defined together with 
the Grassmann parameter $\xi_{A}$'s, 
one needs to take care about the ordering of the component fields
in showing the SUSY invariance,
even though each component field is defined as an ordinary 
bosonic or fermionic function. For instance,
\begin{eqnarray}
\delta_{A}(\phi(x)\tilde{\phi}(x))
\neq \delta_{A}(\tilde{\phi}(x)\phi(x)),
\end{eqnarray}
even though we have $\phi(x)\tilde{\phi}(x) = \tilde{\phi}(x)\phi(x)$.
These properties of the lattice SUSY transformation laws 
essentially necessitates
the notion of ``proper" ordering in the component-wise expressions.
For this reason, we respected the orderings of the component fields
in the expressions (\ref{kinetic_terms})-(\ref{int_terms}).
It is a straightforward task to verify that these proper ordered component-wise
expressions for the action (\ref{kinetic_terms})-(\ref{int_terms})
are manifestly invariant under all the $\cN=D=2$ twisted SUSY transformations
on the lattice.

We note here that the notion of ``proper" ordering 
in the component-wise expression
has been already argued in Ref. \cite{DKKN3,ADKS}
in replying to the critiques claimed in Ref \cite{Bruckmann},
and now we are revisiting this issue again with 
the star product formulation at hand.
It is obvious that the necessity of ``proper" ordering stems from the
intrinsic non-commutative nature of the star product defined in 
(\ref{star_definition})
and is ultimately originated from the nature of the Leibniz rule
for the difference operations. 
Now the question is how this kind of ``proper" ordering can be survived
or respected in the quantum treatment of the formulation.
This is actually the main subject of this paper. 
We anticipate the result and claim here that
respecting the ``proper" ordering 
essentially corresponds to picking up the
planar diagrams in the perturbative expansions. 
Developing the superfield formulation,
we will explicitly see in Sec. 5 
that in the planar diagrams
all the phase factors which stem from the non-commutativity
strictly protect the SUSY structure on the lattice and give the exact SUSY result.
In other words, the above notion of ``proper" ordering 
can be preserved and respected in the planar limit or 
in the `t Hooft large-$N$ limit.
Thus, the exact SUSY w.r.t. all the supercharges
at non-zero lattice spacing is achieved in this limit.
 
We also note some important properties of superfield integrations
under the $x$-summation (\ref{x_sum}).
For the kinetic term or the D-term, we have, 
\begin{eqnarray}
\sum_{x}\int d^{4}\theta\ \oPhi * \Phi
&=& \sum_{x}\int d^{4}\theta\ \oPhi \Phi 
\ =\ \sum_{x}\int d^{4}\theta\ \Phi \oPhi,
\label{Dterm}
\end{eqnarray}
which stems from the vanishing sum of the shift parameters (\ref{a_vanishing_sum}).
For the F-terms in particular for the mass terms, we have,
\begin{eqnarray}
\sum_{x}\int d^{2}\theta\ \Phi_{1} * \Phi_{2}
&=& \sum_{x}\int d^{2}\theta\ \Phi_{1} \Phi_{2}
\ =\ \sum_{x}\int d^{2}\theta\ \Phi_{2} \Phi_{1},  \label{mass1} \\[2pt] 
\sum_{x}\int d^{2}\otheta\ \oPhi_{1} * \oPhi_{2}
&=& \sum_{x}\int d^{2}\otheta\ \oPhi_{1} \oPhi_{2}
\ =\ \sum_{x}\int d^{2}\otheta\ \oPhi_{2} \oPhi_{1},  \label{mass2}
\end{eqnarray}
thanks to the symmetric choice of the shift parameter 
(\ref{a_symmetric_choice}).
Namely, the kinetic and mass terms can be written
without the star product under the $x$-summation (\ref{x_sum}).
This is a preferable feature 
for the superfield formulation,
particularly in deriving the superfield propagators
as we will see in the next subsection.
For the interaction terms, 
we have the cyclic permutation property for the (anti-)chiral superfields,
\begin{eqnarray}
\sum_{x}\int d^{2}\theta\ \Phi_{1} * \Phi_{2} * \Phi_{3} 
&=& \sum_{x}\int d^{2}\theta\ \Phi_{2} * \Phi_{3} * \Phi_{1}
\ =\ \sum_{x}\int d^{2}\theta\ \Phi_{3} * \Phi_{1} * \Phi_{2}, 
\label{int1} \\[2pt]
\sum_{x}\int d^{2}\overline{\theta}\ \oPhi_{1} * \oPhi_{2} * \oPhi_{3} 
&=& \sum_{x}\int d^{2}\overline{\theta}\ \oPhi_{2} * \oPhi_{3} * \oPhi_{1}
\ =\ \sum_{x}\int d^{2}\overline{\theta}\ \oPhi_{3} * \oPhi_{1} * \oPhi_{2},
\label{int2}
\end{eqnarray}
which is again thanks to the symmetric choice (\ref{a_symmetric_choice}).
Actually, one can show that 
the equalities in (\ref{mass1}) and (\ref{int1})
hold only if one takes $a_{1}+a_{2}=0$.
Likewise, 
the equalities in (\ref{mass2}) and (\ref{int2})
hold only if one takes $a + \ta=0$.
This technical advantage 
is the main reason why we take the symmetric choice 
(\ref{a_symmetric_choice}).

\begin{figure}
\begin{center}
\includegraphics[width=90mm]{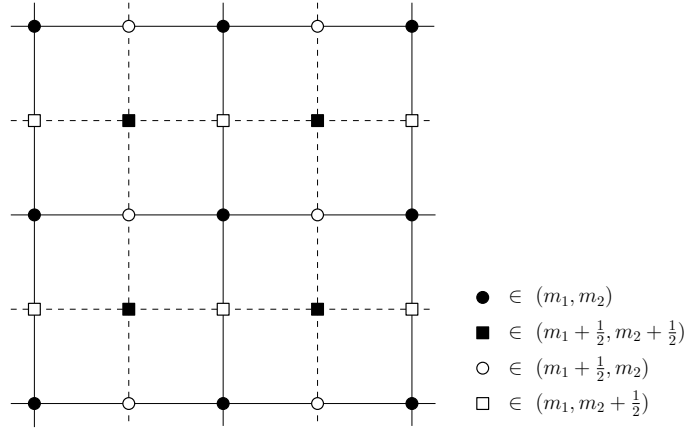}
\caption{Configuration of the superfields in the action
subject to the $x$-summation (\ref{x_sum2})}
\label{fig_superfields_config}
\end{center}
\end{figure}

Although the summation defined in (\ref{x_sum})
is already sufficient for ensuring the SUSY invariant nature of the 
total action, 
in the following for the technical convenience
in the momentum analysis, 
we will conventionally take the $x$-summation to
cover (int, half-int) and (half-int,int) sites
as well,
\begin{eqnarray}
\sum_{x} &=& \sum_{(m_{1},m_{2})}
+ \sum_{(m_{1}+\frac{1}{2},m_{2}+\frac{1}{2})}
+ \sum_{(m_{1}+\frac{1}{2},m_{2})}
+ \sum_{(m_{1},m_{2}+\frac{1}{2})}, \hspace{10pt}
(m_{1},m_{2}\ : \mathrm{integers}).
\label{x_sum2}
\end{eqnarray}
Namely, we consider a finer lattice
which can be measured by half of the unit length.
Accordingly, the field propagations are
now also allowed from (int, half-int) to (int, half-int)
and (half-int, int) to (half-int, int) sites. 
Notice that the lattice action defined by the $x$-summation over
(int,int) and (half-int,half-int) 
is totally independent from the one 
given by (half-int,int) and (int,half-int)
in the sense that they are not correlated each other neither
dynamically nor supersymmetrically.
The multiplicity of the superfields introduced so far is
summarized as follows.
The SUSY invariance essentially requires
the (super)fields defined on both (int,int) and
(half-int, half-int) sites as in (\ref{x_sum}).
This causes the field multiplicity of factor 2.
Furthermore, as explained above, 
we conventionally introduced a trivial multiplicity of 
factor 2 which stems from the summation (\ref{x_sum2}).
The entire multiplicity of the superfields originated from the $x$-summation
is then 
\begin{eqnarray}
2(\mathrm{SUSY}) \times 2(\mathrm{conventional}) = 4.
\label{multiplicity}
\end{eqnarray}
We observe that 
the lattice action subject to the $x$-summation given in
(\ref{x_sum2})
describes four copies of the $D=\cN=2$ twisted Wess-Zumino model
in the naive continuum limit.
We will come back this point when dealing with the momentum space analysis
of the superfield propagators in the next section.
The lattice superfield configuration subject to 
the $x$-summation (\ref{x_sum2}) 
is depicted in Fig. \ref{fig_superfields_config}.

\subsection{Superfield propagators}

\indent

Having defined the $\cN=D=2$ twisted SUSY invariant action on the lattice, 
we shall proceed to set up the technical ingredients
requisite for studying the possible quantum corrections.
We shall take a full advantage
of the lattice superfield method
and intend to calculate the radiative corrections 
entirely in the superfield method.
In what follows we derive the superfield propagators on the lattice.
The notion of superfield propagators
was first introduced in \cite{Fujikawa-Lang,Ferrara-Piguet}.
The derivation in this section basically follows the same procedure 
as in the continuum spacetime \cite{Ferrara-Piguet,Piguet-Sibold} 
or in the Moyal non-commutative spacetime \cite{Bichl}. 
We start from the free part of the tree level vertex functional,
which we shall denote as $\Gamma_{0}$.
It is given by the free part of the lattice action (\ref{total_action}).
Using the relations (\ref{phi_phi_prime}),
we express it 
in terms of $\Phi'$ and $\oPhi'$ in (\ref{phi_prime1})-(\ref{phi_prime2}),
\begin{eqnarray}
\Gamma_{0}[\Phi,\oPhi] &=& 
\sum_{x} \int d^{4}\theta\ 
(U^{2} * \oPhi' * U^{-2}) \Phi'
+\frac{m}{2}\sum_{x} \biggl[
\int d^{2}\theta\ \Phi'\Phi' 
+\int d^{2}\overline{\theta}\ \oPhi'\oPhi' 
\biggr] \label{Gamma1}  \\[2pt]
&=& \sum_{x} \int d^{4}\theta\ 
\oPhi' (U^{-2} * \Phi' * U^{2}) 
+\frac{m}{2}\sum_{x} \biggl[
\int d^{2}\theta\ \Phi'\Phi' 
+\int d^{2}\overline{\theta}\ \oPhi'\oPhi' 
\biggr], \label{Gamma2}
\end{eqnarray} 
where the $U$ is denoting the difference expansion operator
defined in (\ref{U_def}), and
the Grassmann integration measures are again defined by
$d^{4}\theta \equiv d\theta  d\ttheta d\theta_{1}d\theta_{2}$,
$d^{2}\theta \equiv d\theta_{2}d\theta_{1}$ and
$d^{2}\otheta \equiv d\ttheta d\theta$.
The corresponding connected Green's functional,
which we denote as $W_{0}$, is defined by
the Legendre transformation,
\begin{eqnarray}
W_{0}[J,\oJ] &=& \Gamma_{0}[\Phi,\oPhi] 
+ \sum_{x} \biggl[
\int d^{2}\theta\ 
J'\Phi' + \int d^{2}\overline{\theta}\ 
\oJ' \oPhi'
\biggr],  \label{legendre}
\end{eqnarray}
where $J'$ and $\oJ'$ denote the chiral and anti-chiral
source superfields which obey
the same conditions as $\Phi'$ (\ref{Chiral_cond_modified1})
and $\oPhi'$ (\ref{Chiral_cond_modified2}),  
respectively,
\begin{eqnarray}
[\xi \frac{\partial}{\partial\theta}, J'(x,\theta_{A})]_{*} &=& 
[\tilde{\xi}\frac{\partial}
{\partial\tilde{\theta}}, J'(x,\theta_{A})]_{*} \ =\ 0, \\[2pt]
[\xi_{\mu}\frac{\partial}{\partial\theta_{\mu}} ,
\oJ'(x,\otheta_{A})]_{*} &=& 0, \hspace{20pt}
(\mathrm{\mu\ : \ no\ sum}).
\end{eqnarray}
The Legendre transformation (\ref{legendre}) gives rise to 
the relations between
$(J',\oJ')$ and $(\Phi',\oPhi')$,
\begin{eqnarray}
\frac{\delta\Gamma_{0}}{\delta\Phi'} &=& -J' , \hspace{20pt}
\frac{\delta\Gamma_{0}}{\delta\oPhi'} \ =\ -\oJ',  
\label{Legendre1}
\\[4pt]
\frac{\delta W_{0}}{\delta J'} &=& \Phi', \hspace{25pt}
\frac{\delta W_{0}}{\delta \oJ'} \ =\ \oPhi'. 
\label{Legendre2}
\end{eqnarray}
The superfield propagators for $\Phi'$ and $\oPhi'$ are 
defined in terms of the functional derivatives of 
the connected Green's functional w.r.t. the source superfields
$(J',\oJ')$,
\begin{eqnarray}
< \Phi'(1)\Phi'(2)> &\equiv& 
\frac{\delta^{2}}{\delta J'(1) \delta J'(2)} W_{0}[J,\oJ] 
\ =\  \frac{\delta}{\delta J'(1)} \Phi'[J,\oJ] (2), 
\label{superfield_prop_def1}
\\[2pt]
< \oPhi'(1)\Phi'(2)> &\equiv& 
\frac{\delta^{2}}{\delta \oJ'(1) \delta J'(2)} W_{0}[J,\oJ] 
\ =\  \frac{\delta}{\delta \oJ'(1)} \Phi'[J,\oJ] (2),
\label{superfield_prop_def2}
\end{eqnarray} 
where the arguments 1 and 2 refer to each superspace point.
In the second equalities, we used the relations (\ref{Legendre2}).

In order to calculate the superfield propagators, 
we first need to write down the l.h.s.'s of the relations
(\ref{Legendre1}) explicitly, and then solve them w.r.t. $\Phi'$ and $\oPhi'$.
By functionally differentiating the expressions 
(\ref{Gamma1}) and (\ref{Gamma2}) by $\Phi'$ and $\oPhi'$, respectively, 
we actually have,
\begin{eqnarray}
\int d^{2}\overline{\theta}^{(1)}\ U^{2}(1) * \oPhi'(1) * U^{-2}(1) 
+ m \Phi'(1) &=& -J'(1),  \label{J_relation1} \\[2pt]
\int d^{2}\theta^{(1)}\ U^{-2}(1) * \Phi'(1) * U^{2}(1) 
+ m \Phi'(1) &=& -\oJ'(1). \label{J_relation2}
\end{eqnarray}
In deriving them, we made use of the functional differential
relations for $(\Phi',\oPhi')$,
\begin{eqnarray}
\frac{\delta \Phi'(2)}{\delta \Phi'(1)} 
&=& \delta^{2}_{x^{(1)},x^{(2)}}\delta^{2}(\theta^{(1)}-\theta^{(2)}),
\hspace{20pt}
\frac{\delta \oPhi'(2)}{\delta \oPhi'(1)} 
\ =\ \delta^{2}_{x^{(1)},x^{(2)}}\delta^{2}(\otheta^{(1)}-\otheta^{(2)}),
\label{phi_derivative}
\end{eqnarray}
where $\delta^{2}_{x^{(1)},x^{(2)}}$ denotes the two dimensional 
Kronecker delta,
while the delta functions for the Grassmann coordinates
$\delta^{2}(\theta^{(1)}-\theta^{(2)})$ and 
$\delta^{2}(\otheta^{(1)}-\otheta^{(2)})$
are defined by
\begin{eqnarray}
\delta^{2}(\theta^{(1)}-\theta^{(2)}) &=& 
\prod_{A}(\theta^{(1)}_{A}-\theta^{(2)}_{A})
\ =\ (\theta^{(1)}_{1}-\theta^{(2)}_{1})
(\theta^{(1)}_{2}-\theta^{(2)}_{2}), \\[2pt]
\delta^{2}(\otheta^{(1)}-\otheta^{(2)}) &=& 
\prod_{A}(\otheta^{(1)}_{A}-\otheta^{(2)}_{A})
\ =\ (\theta^{(1)}-\theta^{(2)})
(\ttheta^{(1)}-\ttheta^{(2)}).
\end{eqnarray} 
Operating $U^{2}\, *\,$ from the left and $\,*\, U^{-2}$ from the right on 
(\ref{J_relation2}) and taking the $d^{2}\overline{\theta}$ integration
would give
\begin{eqnarray}
\Delta_{\mu}^{2}
\Phi'(1)
+ m\int d^{2}\overline{\theta}^{(1)}\ U^{2}(1) * \oPhi'(1) * U^{-2}(1)  
= 
-\int d^{2}\overline{\theta}^{(1)}\ U^{2}(1) * \oJ'(1) * U^{-2}(1),
\label{J_relation3} 
\end{eqnarray}
and the similar expression for (\ref{J_relation1})
by operating $U^{-2}\, *\,$ from the left and $\, *\, U^{2}$ from the right.
In (\ref{J_relation3}) 
we used the following relation which can be verified explicitly,
\begin{eqnarray}
\int d^{2}\overline{\theta}\ U^{2} *
\biggl[
\int d^{2}\theta\ U^{-2} * \Phi'(x,\theta_{A}) * U^{2}
\biggr]
* U^{-2} 
= \Delta_{\mu}^{2}
\Phi'(x,\theta_{A}). 
\end{eqnarray}
By substituting (\ref{J_relation1}) into (\ref{J_relation3})
and solving it w.r.t. $\Phi'$ we then obtain
\begin{eqnarray}
\Phi'(1) &=& \frac{-1}{\Delta_{\mu}^{2}-m^{2}}
\int d^{2}\overline{\theta}^{(1)}\ U^{2}(1)*\oJ'(1) * U^{-2}(1)
+\frac{m}{\Delta_{\mu}^{2}-m^{2}}J'(1).
\label{Phi_relation1}
\end{eqnarray}
Likewise we obtain for $\oPhi'$,
\begin{eqnarray}
\oPhi'(1) &=& \frac{-1}{\Delta_{\mu}^{2}-m^{2}}
\int d^{2}\theta^{(1)} \ U^{-2}(1)*J'(1) * U^{2}(1)
+\frac{m}{\Delta_{\mu}^{2}-m^{2}}\oJ'(1).
\label{Phi_relation2}
\end{eqnarray}
By functionally differentiating the relations 
(\ref{Phi_relation1})-(\ref{Phi_relation2}) w.r.t. the source superfields 
$J'$ and $\oJ'$
and using the definitions 
(\ref{superfield_prop_def1})-(\ref{superfield_prop_def2}),
we finally obtain the superfield propagators which are given by
\begin{eqnarray}
< \Phi'(x^{(1)},\theta^{(1)}_{A})\Phi'(x^{(2)},\theta^{(2)}_{A})> 
&=& \frac{m}{\Delta_{\mu}^{2}-m^{2}}
\delta^{2}_{x^{(1)},x^{(2)}}
\delta^{2}(\theta^{(1)}-\theta^{(2)}) 
\nonumber \\[2pt]
&=& \int^{2\pi}_{-2\pi} \frac{d^{2}p}{(4\pi)^{2}}  
e^{ip(x^{(1)}-x^{(2)})}
\frac{-m}{\sin^{2}{p_{\mu}}+m^{2}}
\delta^{2}(\theta^{(1)}-\theta^{(2)}), \qquad \label{phiphi_prop1}
\\[2pt]
< \oPhi'(x^{(1)},\otheta^{(1)}_{A})\oPhi'(x^{(2)},\otheta^{(2)}_{A})> 
&=& \frac{m}{\Delta_{\mu}^{2}-m^{2}}
\delta^{2}_{x^{(1)},x^{(2)}}
\delta^{2}(\otheta^{(1)}-\otheta^{(2)})  
\nonumber \\[2pt]
&=& \int^{2\pi}_{-2\pi} \frac{d^{2}p}{(4\pi)^{2}}  
e^{ip(x^{(1)}-x^{(2)})}
\frac{-m}{\sin^{2}{p_{\mu}}+m^{2}}
\delta^{2}(\otheta^{(1)}-\otheta^{(2)}), \\[8pt] 
\label{phiphi_prop2}
< \oPhi'(x^{(1)},\otheta^{(1)}_{A})\Phi'(x^{(2)},\theta^{(2)}_{A})> 
&=& \frac{-1}{\Delta_{\mu}^{2}-m^{2}}
e^{+i E^{(12)}_{\mu}\Delta_{\mu}^{(1)}} 
\ \delta^{2}_{x^{(1)},x^{(2)}} 
\nonumber  \\[2pt]
&=& \int^{2\pi}_{-2\pi}\frac{d^{2}p}{(4\pi)^{2}}
e^{ip(x^{(1)}-x^{(2)})}
\frac{e^{-E^{(12)}_{\mu}\sin{p_{\mu}}}}
{\sin^{2}{p_{\mu}}+m^{2}},  \\[2pt]
< \Phi'(x^{(1)},\theta^{(1)}_{A})\oPhi'(x^{(2)},\otheta^{(2)}_{A})> 
&=& \frac{-1}{\Delta_{\mu}^{2}-m^{2}}
e^{-i E^{(21)}_{\mu}\Delta_{\mu}^{(1)}} 
\ \delta^{2}_{x^{(1)},x^{(2)}} 
\nonumber \\[2pt]
&=& \int^{2\pi}_{-2\pi}\frac{d^{2}p}{(4\pi)^{2}}
e^{ip(x^{(1)}-x^{(2)})}
\frac{e^{+E^{(21)}_{\mu}\sin{p_{\mu}}}}
{\sin^{2}{p_{\mu}}+m^{2}},  \label{phiphi_prop4}
\end{eqnarray}
where the symbols $E^{(12)}$ and $E^{(21)}$ are defined by
$E^{(ij)}_{\mu}\equiv\theta^{(i)}\theta_{\mu}^{(j)}
+\epsilon_{\mu\nu}\tilde{\theta}^{(i)}\theta_{\nu}^{(j)}$ 
for $i,j=1\ \mathrm{or}\ 2$.
In each second line of these expressions, 
we inserted the momentum space representation of $\delta^{2}_{x^{(1)},x^{(2)}}$,
\begin{eqnarray}
\delta^{2}_{x^{(1)},x^{(2)}}
&=& \int^{2\pi}_{-2\pi} \frac{d^{2}p}{(2\pi)^{2}}
e^{ip(x^{(1)}-x^{(2)})}.
\end{eqnarray}
Notice that the 
the range of the Brillouin zone is $4\pi$ for each direction
since we introduced the superfields located on
the half-integer sites for each spacetime direction.
Within that range of the Brillouin zone,
each propagator (\ref{phiphi_prop1})-(\ref{phiphi_prop4}) 
has superficially 16 spectrum doubling,
namely, 4 species for each direction.

It is important to recognize that the 
entire doubling is originated from the two reasons.
The first one
is due to the $x$-summation over the half-integer sites
(\ref{x_sum2}).
This can be regarded as a trivial multiplicity
as far as the superfield propagators are concerned,
since the field propagations are only restricted
from 
(int,int) to (int,int),
(int,half-int) to (int,half-int),
(half-int,int) to (half-int,int)
and (half-int,half-int) to (half-int,half-int)
sites.
In the configuration space, this is clear from the fact that
each term in the kinetic action (\ref{kinetic_terms})
consists of (int,int), (half-int,half-int), (int,half-int) or (half-int,int)
sites.
In the momentum space this feature 
is reflected to the 
$2\pi$ periodicity of the integrand
in each propagator except for the exponential factor.
In fact, starting from the above expressions
of the propagators (\ref{phiphi_prop1})-(\ref{phiphi_prop4}),
one easily sees that there are no correlations between 
the integer and the half-integer sites
for each direction.
For instance, the propagator (\ref{phiphi_prop1}) 
can be re-expressed as
\renewcommand{\arraystretch}{1.5}
\begin{eqnarray}
&&< \Phi'(x^{(1)},\theta^{(1)}_{A})\Phi'(x^{(2)},\theta^{(2)}_{A})> 
\nonumber  \\[4pt]
&& \hspace{5pt}
= \biggl\{
\begin{array}{l}
\displaystyle
\int^{\pi}_{-\pi}\frac{d^{2}p}{(2\pi)^{2}}
e^{ip(x^{(1)}-x^{(2)})}
\frac{-m}{\sin^{2}{p_{\mu}}+m^{2}}
\delta^{2}(\theta^{(1)}-\theta^{(2)}), \ \ 
\mathrm{for} \ \ x^{(1)}-x^{(2)} = (m_{1},m_{2}) \qquad \\
0, \ \ \ \mathrm{otherwise}
\end{array}
\end{eqnarray}
where $m_{1}$ and $m_{2}$ denote any integer values.
Note that the range of the Brillouin zone
for the non-vanishing propagator is now taken as $2\pi$
since the integrand is actually $2\pi$ periodic.
In deriving this expression, we  divided the 
momentum regions into two of $2\pi$ and
made use of the $2\pi$ (anti-)periodicity
of the exponential factor $e^{ip(x^{(1)}-x^{(2)})}$.
The other propagators 
(\ref{phiphi_prop2})-(\ref{phiphi_prop4})
also have the same feature.
This observation implies that the factor $4$ in the entire multiplicity
due to the $x$-summation over half-integer sites
are just corresponding to the four copies of the 
physically equivalent momentum regions
as long as the propagators are concerned.

In contrast,
the rest of the spectrum doubling which 
still remains in the propagators
has a non-trivial physical meaning 
via the $D=\cN=2$ Dirac-K\"ahler mechanism.
This stems from the fact that
the chiral and anti-chiral superfields 
$(\Phi',\oPhi')$ contain the components of 
the Dirac-K\"ahler twisted fermions 
$(\chi,\psi_{\mu},\tchi)$ which turn into the 
corresponding staggered fermion components on the lattice.
In this procedure, the extended $\cN=2$ SUSY index can be regarded
as $N_{f}=2$ flavor (taste) index via the 
Dirac-K\"ahler expansion of the component fields on the lattice,
\begin{eqnarray}
\xi_{\alpha i}(x)
&=&
(\mathbf{1}\chi(x)+\gamma_{\mu}\psi_{\mu}(x+n_{\mu})
+\gamma_{5}\tchi(x+n_{1}+n_{2}))_{\alpha i},
\label{xi}
\end{eqnarray}  
where the suffices $\alpha=1,2$ and $i=1,2$ are representing
the spinor and the internal indices of $\cN=2$ twisted SUSY.
This procedure has been already pointed out in \cite{DKKN1}
and it resolves the multiplicity of 4 
into $2(\mathrm{spinor}) \times 2(\cN=2)$.
In order to see how this mechanism is incorporated in the 
present superfield formulation, it is instructive to
derive the propagators for the component fields
in terms of the $\xi_{\alpha i}$ basis which is given by 
\begin{eqnarray}
<\xi_{\alpha i}(x^{(1)})\overline{\xi}_{j\beta}(x^{(2)})>
&=&
<(\mathbf{1}\chi(x^{(1)})+\gamma_{\mu}\psi_{\mu}(x^{(1)}+n_{\mu})
+\gamma_{5}\tchi(x^{(1)}+n_{1}+n_{2}))_{\alpha i} \nonumber \\[2pt]
&&\times 
(\mathbf{1}\chi(x^{(2)})+\gamma_{\mu}\psi_{\mu}(x^{(2)}+n_{\mu})
-\gamma_{5}\tchi(x^{(2)}+n_{1}+n_{2}))_{j \beta} >, \qquad
\label{prop_xi_def}
\end{eqnarray}
where $\overline{\xi}$ is defined by the transpose of $\xi$, 
$\overline{\xi}\equiv\xi^{T}$,
and the $\gamma_{\mu}$ matrices are given in (\ref{Pauli_matrices}).
Since $\xi_{\alpha i}$ is made up with the component fields 
located in every corner of the square with one integer lattice spacing 
each side (\ref{xi}),
the locational separation $|x^{(1)}-x^{(2)}|$ 
in terms of $\xi_{\alpha i}$ should be taken
as even number for each direction.
This implies that
the physical range of the Brillouin zone for $\xi_{\alpha i}$ should
be restricted as $\pi$ instead of $2\pi$ for each direction. 
Each correlation function for the component fields can be 
obtained by differentiating the corresponding superfield propagator
w.r.t. $\theta$'s, for instance, 
\begin{eqnarray}
<\chi(x^{(1)})\psi_{\mu}(x^{(2)}+n_{\mu})>
&=& 
\frac{\partial}{\partial \theta^{(1)}}\frac{\partial}{\partial \theta_{\mu}^{(2)}}
<\oPhi'(x^{(1)},\otheta_{A})\Phi'(x^{(2)}+n_{\mu},\theta_{A})> 
|_{\theta_{A}=\otheta_{A}=0}^{|x^{(1)}-x^{(2)}|:even}
\nonumber \\[2pt]
&=& 
\int_{-\frac{\pi}{2}}^{\frac{\pi}{2}} \frac{d^{2}p}{(2\pi)^{2}}
e^{ip(x^{(1)}-x^{(2)})}e^{-ip_{\mu}}
\frac{\sin p_{\mu}}{\sin^{2} p_{\rho}+ m^{2}}.  
\end{eqnarray}
Making the use of the completeness relation for the $\gamma_{\mu}$ matrices,
\begin{eqnarray}
(\mathbf{1})_{ij}(\mathbf{1})_{kl}
+ (\gamma_{\mu})_{ij} (\gamma_{\mu})_{kl}
+ (\gamma_{5})_{ij} (\gamma_{5})_{kl}
&=& 2\delta_{ik}\delta_{jl},
\end{eqnarray}
one obtains the following expression of the 
propagator for the $\xi_{\alpha i}$ basis,
\begin{eqnarray}
<\xi_{\alpha i}(x^{(1)})\overline{\xi}_{j\beta}(x^{(2)})>
&=& 
\int_{-\frac{\pi}{2}}^{\frac{\pi}{2}} \frac{d^{2}p}{(2\pi)^{2}}
e^{ip(x^{(1)}-x^{(2)})}
\frac{1}{\sin^{2} p_{\rho}+ m^{2}}   \nonumber \\
&& \times  \sum_{\mu}
\biggl[
(\gamma_{\mu})_{\alpha\beta}\delta_{ij}\sin p_{\mu}
-2i (\gamma_{5})_{\alpha\beta} (\gamma_{5}\gamma_{\mu})_{ij}
\sin^{2}p_{\mu} -M_{\alpha\beta,ij}
\biggr],  \qquad
\label{prop_xi}
\end{eqnarray}
where the mass matrix $M_{\alpha\beta, ij}$ is given by
\begin{eqnarray}
M_{\alpha\beta, ij} &=&
2m\bigl[
\delta_{\alpha\beta}(\gamma_{5})_{ij}\cos p_{1} \cos p_{2}
+ (\gamma_{5})_{\alpha\beta}\delta_{ij} \sin p_{1} \sin p_{2} 
\nonumber \\[2pt]
&& - i (\gamma_{1})_{\alpha\beta} (\gamma_{2})_{ij} \cos p_{1} \sin p_{2}
+ i (\gamma_{2})_{\alpha\beta} (\gamma_{1})_{ij} \sin p_{1} \cos p_{2}
\bigr].
\end{eqnarray}
The integrand in (\ref{prop_xi})
can essentially be regarded as the momentum space expression of
the two dimensional $N_{f}=2$
staggered fermion propagator in terms of spin-flavor(taste) 
basis.
Notice that because the range of the Brillouin zone is $\pi$
for each direction, we do not have any spectrum doubling in this fermion
basis.
Also note that thanks to the manifest superfield formulation, this mechanism 
also provides a physical interpretation for the spectrum doubling in
the bosonic sector.
To summarize, 
the entire multiplicity which is 16 can be decomposed into,
\begin{eqnarray}
16 &=& 4 (\textrm{Dirac-K\"ahler}) \times 4 (x\textrm{-summation}) \\[2pt]
&=& 2(\textrm{spinor}) \times 2(\cN=2) \times 2(\textrm{SUSY}) \times
2(\textrm{conventional}),
\end{eqnarray}
where, as we have seen, the factor
4 ($x$\textrm{-summation}) can be regarded as
the trivial multiplicity as far as the 
propagators are concerned, while 
for the factor 4 (\textrm{Dirac-K\"ahler})
the Dirac-K\"ahler mechanism plays a physically substantial role. 

\begin{figure}
\begin{center}
\includegraphics[width=70mm]{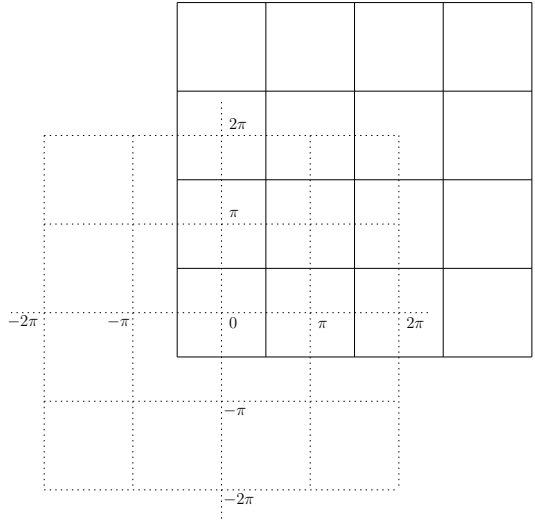}
\caption{16 physical integration regions in momentum space (solid lines)}
\label{fig_BZ}
\end{center}
\end{figure}

Having known the origins of the spectrum doubling,
we divide the entire Brillouin zone,
$-2\pi \leq p_{\mu} \leq 2\pi\ (\mu =1,2)$, 
into $4\times 4=16$ physical momentum regions
by following the similar manner as in the
staggered fermion momentum space analysis \cite{Sharatchandra}
(see Fig. \ref{fig_BZ}),
\begin{eqnarray}
p_{\mu} &=& k_{\mu} +  2(\pi_{K})_{\mu} + (\pi_{L})_{\mu}, 
\label{momentum}
\hspace{20pt}
\mathrm{with} \ \ \
-\frac{\pi}{2} \leq k_{\mu} \leq \frac{\pi}{2},
\label{physical_momentum}
\end{eqnarray}
where the $\pi_{K}$ and $\pi_{L}$ 
respectively  with four indices,
$K,L = 00,10,01,11$,
are defined by
\begin{eqnarray}
\pi_{00} &=& (0,0), \hspace{10pt} 
\pi_{10} \ =\ (\pi,0), \hspace{10pt} 
\pi_{01} \ =\ (0,\pi), \hspace{10pt} 
\pi_{11} \ =\ (\pi,\pi). 
\end{eqnarray}
In (\ref{physical_momentum}), the $2\pi_{K}$ spans the entire momentum
space into four copies of physically equivalent momentum regions
which corresponds to the multiplicity of 4($x$-summation),
while the $\pi_{L}$ further spans each regions into
four smaller regions which corresponds to the multiplicity of
4(Dirac-K\"ahler).
The Fourier transformations of the superfields are
accordingly given by
\begin{eqnarray}
\Phi'(x,\theta_{A})
&=& \int^{2\pi}_{-2\pi}
\frac{d^{2}p}{(4\pi)^{2}}
e^{ipx}\Phi'(p,\theta_{A}) 
\ =\ 
\sum_{K,L} \int^{\frac{\pi}{2}}_{-\frac{\pi}{2}}
\frac{d^{2}k}{(4\pi)^{2}}
e^{i(k+2\pi_{K}+\pi_{L})x}\Phi'_{KL}(k,\theta_{A}), \qquad \\[2pt]
\oPhi'(x,\otheta_{A})
&=& \int^{2\pi}_{-2\pi}
\frac{d^{2}p}{(4\pi)^{2}}
e^{ipx}\oPhi'(p,\otheta_{A})  
\ =\ 
\sum_{K,L} \int^{\frac{\pi}{2}}_{-\frac{\pi}{2}}
\frac{d^{2}k}{(4\pi)^{2}}
e^{i(k+2\pi_{K}+\pi_{L})x}\oPhi'_{KL}(k,\otheta_{A}), 
\end{eqnarray}
where the superfields in the momentum space
with indices $K$ and $L$ are defined by
\begin{eqnarray}
\Phi'_{KL}(k,\theta_{A}) &\equiv& \Phi'(k+2\pi_{K}+\pi_{L},\theta_{A}), \hspace{10pt}
\oPhi'_{KL}(k,\otheta_{A}) \ \equiv\ \oPhi'(k+2\pi_{K}+\pi_{L},\otheta_{A}). 
\end{eqnarray}
In terms of the superfields $(\Phi'_{KL},\oPhi'_{KL})$, 
the propagators are given by
\begin{eqnarray}
< \Phi'_{K^{(1)}L^{(1)}}(k^{(1)},\theta^{(1)}_{A})
\Phi'_{K^{(2)}L^{(2)}}(k^{(2)},\theta^{(2)}_{A})> 
&=& 
(4\pi)^{2}
\delta_{K^{(1)},K^{(2)}}
\delta_{L^{(1)},L^{(2)}}\delta^{2}(k^{(1)}+k^{(2)})  \nonumber \\
&& \times
\frac{-m}{\sin^{2}{k_{\mu}^{(1)}}+m^{2}}
\delta^{2}(\theta^{(1)}-\theta^{(2)}),  \label{k_prop1} \\[2pt]
< \oPhi'_{K^{(1)}L^{(1)}}(k^{(1)},\otheta^{(1)}_{A})
\oPhi'_{K^{(2)}L^{(2)}}(k^{(2)},\otheta^{(2)}_{A})> 
&=& 
(4\pi)^{2}
\delta_{K^{(1)},K^{(2)}}
\delta_{L^{(1)},L^{(2)}}\delta^{2}(k^{(1)}+k^{(2)}) \nonumber \\
&& \times \frac{-m}{\sin^{2}{k_{\mu}^{(1)}}+m^{2}}
\delta^{2}(\otheta^{(1)}-\otheta^{(2)}), 
\end{eqnarray}
\begin{eqnarray}
< \oPhi'_{K^{(1)}L^{(1)}}(k^{(1)},\otheta^{(1)}_{A})
\Phi'_{K^{(2)}L^{(2)}}(k^{(2)},\theta^{(2)}_{A})> 
&=& 
(4\pi)^{2}
\delta_{K^{(1)},K^{(2)}}
\delta_{L^{(1)},L^{(2)}}\delta^{2}(k^{(1)}+k^{(2)}) \nonumber \\
&& \times \frac{e^{-E^{(12)}_{\mu}\sin{(k+2\pi_{K}+\pi_{L})^{(1)}_{\mu}}}}
{\sin^{2}{k_{\mu}^{(1)}}+m^{2}}, \\[2pt]
< \Phi'_{K^{(1)}L^{(1)}}(k^{(1)},\theta^{(1)}_{A})
\oPhi'_{K^{(2)}L^{(2)}}(k^{(2)},\otheta^{(2)}_{A})> 
&=&
(4\pi)^{2}
\delta_{K^{(1)},K^{(2)}}
\delta_{L^{(1)},L^{(2)}}\delta^{2}(k^{(1)}+k^{(2)}) \nonumber \\
&& \times \frac{e^{+E^{(21)}_{\mu}\sin{(k+2\pi_{K}+\pi_{L})^{(1)}_{\mu}}}}
{\sin^{2}{k_{\mu}^{(1)}}+m^{2}}.  \label{k_prop4}
\end{eqnarray}
Since the momentum range for $k$ is defined from $-\frac{\pi}{2}$ to
$\frac{\pi}{2}$,
these propagators (\ref{k_prop1})-(\ref{k_prop4})
do not have any spectrum doubling
in each physical momentum division.
Notice that the propagators are diagonal w.r.t. the indices $K$ and $L$.
namely, each physical momentum division is not mixed up each other
while propagating.
Also note that 
as far as the internal lines are concerned
we are still practically allowed to use the notation $p$
and to integrate over the entire momentum region from $-2\pi$ to $2\pi$,
keeping in mind that all the 16 physical momenta are actually contributing
to each internal line.
This is because the $k$, $\pi_{K}$ and $\pi_{L}$ always come up together
and we eventually sum up all the momentum divisions $\pi_{K}$ and $\pi_{L}$
for each internal momentum $k$.


\subsection{Three-point vertex functions at tree level}

\indent

After deriving the superfield propagators, we then turn to 
consider the superfield three-point vertex functions.
In contrast to the kinetic and mass terms
which can be expressed without using the star product
(\ref{Dterm})-(\ref{mass2}), 
the interaction terms cannot be free from the non-commutativity
stemming from the star product.
Namely, the physical implications originated from the non-commutativity
in this model are 
essentially inherited in the three-point vertex functions.
In what follows we derive the superfield three-point vertex functions
at tree level which we denote
$\Gamma_{\Phi\Phi\Phi}^{(tree)}$ and $\Gamma_{\oPhi\oPhi\oPhi}^{(tree)}$. 
We start from the tree level vertex functional for 
the interaction terms
expressed in terms of $\Phi'$ and $\oPhi'$ in
(\ref{phi_prime1})-(\ref{phi_prime2}).
It is given by,
\begin{eqnarray}
\Gamma_{int} &=& \frac{g}{3!}\sum_{x}
\biggl[
\int d^{2}\theta\
\Phi'*\Phi'*\Phi'
+\int d^{2}\otheta\
\oPhi'*\oPhi'*\oPhi'
\biggr]. \label{vertex_functional}
\end{eqnarray}
In order to properly take into account 
the effect of the star product, 
it is the most appropriate to Fourier transform the superfields 
not only to the ordinary momentum space $p_{\mu}$ but also 
to the momentum counterpart of the
Grassmann coordinates $\theta_{A}$ and $\otheta_{A}$
which we shall denote as $\chi_{A}=(\chi_{1},\chi_{2})$
and $\ochi_{A}=(\chi,\tchi)$,
\begin{eqnarray}
\Phi'(x,\theta_{A}) &=& \int dp 
\int d^{2}\chi\ e^{ipx}e^{\chi_{A}\theta_{A}}\Phi'(p,\chi_{A}), 
\label{Phi_momentum1} \\[2pt]
\oPhi'(x,\otheta_{A}) &=& \int dp 
\int d^{2}\ochi\ e^{ipx}e^{\ochi_{A}\otheta_{A}}\oPhi'(p,\ochi_{A}), 
\label{Phi_momentum2}
\end{eqnarray} 
where we used the abbreviations for notational simplicity,
$\int dp \equiv \int ^{2\pi}_{-2\pi}\frac{d^{2}p}{(4\pi)^{2}}$,
$\int d^{2}\chi \equiv \int d\chi_{2}d\chi_{1}$
and $\int d^{2}\ochi \equiv \int d\tchi d\chi$
while $\chi_{A}\theta_{A}\equiv \chi_{1}\theta_{1}+\chi_{2}\theta_{2}$
and $\ochi_{A}\otheta_{A}\ \equiv \chi\theta + \tchi\ttheta$.
It is understood that the momentum $p_{\mu}$
should be decomposed into the physical momenta $k_{\mu}$ 
via the relation (\ref{momentum})
whenever necessary.
In terms of the expansions (\ref{Phi_momentum1}) and (\ref{Phi_momentum2}),
we have for the three-point interaction term in the $\Phi'$ sector,
\begin{eqnarray}
&&\hspace{-25pt}\sum_{x} \int d^{2}\theta \
\Phi'(x,\theta_{A})*\Phi'(x,\theta_{A})*\Phi'(x,\theta_{A})
\nonumber \\
&=& \sum_{x} \int d^{2}\theta 
\int \prod_{i=1}^{3}dp^{(i)}d^{2}\chi^{(i)} 
\ 
e^{ip^{(1)}x}e^{\chi^{(1)}_{A}\theta_{A}}*
e^{ip^{(2)}x}e^{\chi^{(2)}_{A}\theta_{A}}*
e^{ip^{(3)}x}e^{\chi^{(3)}_{A}\theta_{A}} \ \nonumber \\[6pt]
&&\times
\Phi'(p^{(1)},\chi^{(1)}_{A})
\Phi'(p^{(2)},\chi^{(2)}_{A})
\Phi'(p^{(3)},\chi^{(3)}_{A}), 
\label{Phi3_exp} \qquad
\end{eqnarray}
and the similar expression for the $\oPhi'$ sector.
The exponential factors in the r.h.s. of (\ref{Phi3_exp})
can be evaluated by using the
definition of the star product (\ref{star_definition}),
\begin{eqnarray}
&&\hspace{-45pt}
e^{ip^{(1)}x}e^{\chi^{(1)}_{A}\theta_{A}}*
e^{ip^{(2)}x}e^{\chi^{(2)}_{A}\theta_{A}}*
e^{ip^{(3)}x}e^{\chi^{(3)}_{A}\theta_{A}} \nonumber \\[2pt]
&=& 
e^{(
\chi_{A}^{(1)-p^{(2)}-p^{(3)}}
+\chi_{A}^{(2)p^{(1)}-p^{(3)}}
+\chi_{A}^{(3)p^{(1)}+p^{(2)}})
\theta_{A}} \
e^{i(p^{(1)}+p^{(2)}+p^{(3)})x},
\end{eqnarray}
where in the r.h.s. we introduced the notations for the ``dressed" $\chi_{A}$'s 
defined by, for instance,
\begin{eqnarray}
\chi_{A}^{(3)p^{(1)}+p^{(2)}} \equiv \chi_{A}^{(3)}e^{ia_{A}(p^{(1)}+p^{(2)})}
=\chi_{A}^{(3)}e^{i(a_{A})_{\mu}(p^{(1)}+p^{(2)})_{\mu}},\ \ 
\  (A : \rm{no\ sum}), \ \ \rm{etc..}
\label{dressed_chi}
\end{eqnarray} 
The $a_{A}$ is denoting $a_{A}=(a_{1},a_{2})$ for the $\Phi'$ sector.
It is clear that the phase factors in the ``dressed" form of the 
Grassmann variables
are originated from the non-commutativity in the configuration space.
After integrating $d^{2}\theta$ and summing over $x$
and then putting back the inverse Fourier transformations,
\begin{eqnarray}
\Phi'(p^{(i)},\chi^{(i)}_{A}) &=&
\sum_{x^{(i)}} \int d^{2}\theta^{(i)}
e^{-ip^{(i)}x^{(i)}}e^{-\chi^{(i)}_{A}\theta^{(i)}_{A}}
\Phi'(x^{(i)},\theta^{(i)}_{A}),
\end{eqnarray}
we obtain,
\begin{eqnarray}
&&\hspace{-25pt}\sum_{x} \int d^{2}\theta \
\Phi'(x,\theta_{A})*\Phi'(x,\theta_{A})*\Phi'(x,\theta_{A})
\nonumber \\
&=& \sum_{x^{(1)}}\sum_{x^{(2)}}\sum_{x^{(3)}} 
\int
\prod_{i=1}^{3} d^{2}\theta^{(i)} dp^{(i)}
\ e^{-i(p^{(1)}x^{(1)}+p^{(2)}x^{(2)}+p^{(3)}x^{(3)})}
\ \delta^{2}_{P}(p^{(1)}+p^{(2)}+p^{(3)}) \nonumber \\[6pt]
&& \times\ \prod_{A}(\theta_{A}^{(1)-p^{(3)}}-\theta_{A}^{(2)})
(\theta_{A}^{(1)p^{(2)}}-\theta_{A}^{(3)}) 
\ \Phi(x^{(1)},\theta^{(1)}_{A})
\Phi(x^{(2)},\theta^{(2)}_{A})
\Phi(x^{(3)},\theta^{(3)}_{A}), \qquad
\label{PhiPhiPhi}
\end{eqnarray}
where 
we introduced the notations for 
the ``dressed" $\theta_{A}$'s
in a similar way as in (\ref{dressed_chi}),
\begin{eqnarray}
\theta_{A}^{(1)p^{(2)}} \equiv 
\theta^{(1)}_{A} e^{ia_{A}p^{(2)}} 
=\theta^{(1)}_{A}e^{i(a_{A})_{\mu}p^{(2)}_{\mu}}, \ \ \ 
(A : \rm{no\ sum}) \ \ \rm{etc..}
\label{dressed_theta}
\end{eqnarray}

We shall give several technical remarks regarding the expressions
(\ref{PhiPhiPhi}) and (\ref{dressed_theta}).
The symbol $\delta^{2}_{P}(...)$ in (\ref{PhiPhiPhi})
is denoting a two-dimensional periodic delta function
mod $4\pi$ for each direction
which obviously stems from the $x$-summation over integer and half-integer sites.
Compatibly,
the phase factors in the ``dressed" $\theta_{A}$'s 
(e.g. (\ref{dressed_theta})) have also $4\pi$ periodicity for each direction.
This is clear from the fact that in the symmetric choice
each $a_{A}$ takes the value of 
$\pm \frac{1}{2}$ for each direction (\ref{a_symmetric_choice}).
The $\theta_{A}$ factors in (\ref{PhiPhiPhi}),
\begin{eqnarray}
\prod_{A}(\theta^{(1)-p^{(3)}}_{A}-\theta^{(2)}_{A}) 
\ \ \ \ \mathrm{and} \ \ \ \
\prod_{A}(\theta^{(1)p^{(2)}}_{A}-\theta^{(3)}_{A})
\end{eqnarray}
are describing the delta functions for the Grassmann variables
with the supports,
\begin{eqnarray}
\theta^{(1)}_{1}e^{-i(a_{1})_{\mu}p^{(3)}_{\mu}}
= \theta^{(2)}_{1},\ \ 
\theta^{(1)}_{2}e^{-i(a_{2})_{\mu}p^{(3)}_{\mu}}
= \theta^{(2)}_{2}, \ \  
\mathrm{and} \ \ \  
\theta^{(1)}_{1}e^{i(a_{1})_{\mu}p^{(2)}_{\mu}}
= \theta^{(3)}_{1},\ \
\theta^{(1)}_{2}e^{i(a_{2})_{\mu}p^{(2)}_{\mu}}
= \theta^{(3)}_{2},
\nonumber 
\end{eqnarray}
respectively.
It is worthwhile to note that each $\theta_{A}$ factor
has the following convenient nature thanks to 
the symmetric choice, $a_{1}+a_{2}=0$,
\begin{eqnarray}
\prod_{A}(\theta^{(1)-p^{(3)}}_{A}-\theta^{(2)}_{A}) &=&
(\theta^{(1)}_{1}e^{-i(a_{1})_{\mu}p^{(3)}_{\mu}}
- \theta^{(2)}_{1})
(\theta^{(1)}_{2}e^{-i(a_{2})_{\mu}p^{(3)}_{\mu}}
- \theta^{(2)}_{2})  \nonumber \\[-4pt]
&=& 
e^{-i(a_{1}+a_{2})_{\mu}p^{(3)}_{\mu}}
(\theta^{(1)}_{1}
- \theta^{(2)}_{1}e^{i(a_{1})_{\mu}p^{(3)}_{\mu}})
(\theta^{(1)}_{2}
- \theta^{(2)}_{2}e^{i(a_{2})_{\mu}p^{(3)}_{\mu}}) \nonumber \\[4pt]
&=&
\prod_{A}(\theta^{(1)}_{A}-\theta^{(2)p^{(3)}}_{A}), 
\label{theta_prod_property}
\end{eqnarray} 
and in a similar way, we have 
$\prod_{A}(\theta^{(1)p^{(2)}}_{A}-\theta^{(3)}_{A})=
\prod_{A}(\theta^{(1)}_{A}-\theta^{(3)-p^{(2)}}_{A})$.
One can use the above properties of the $\theta_{A}$ factors
together with the momentum conservation mod $4\pi$
wherever they are convenient or necessary in the intermediate calculations.

Let us now derive the three-point vertex functions 
$\Gamma_{\Phi\Phi\Phi}^{(tree)}$ and $\Gamma_{\oPhi\oPhi\oPhi}^{(tree)}$
by functionally differentiating the vertex functional (\ref{vertex_functional})
w.r.t. $\Phi'$ and $\oPhi'$, respectively.
Using the expression (\ref{PhiPhiPhi}), 
we  obtain
\begin{eqnarray}
&&\hspace{-20pt}\Gamma_{\Phi\Phi\Phi}^{(tree)}
(x^{(1)},\theta_{A}^{(1)};x^{(2)},\theta_{A}^{(2)};x^{(3)},\theta_{A}^{(3)})
\nonumber \\[2pt]
&&= \frac{\delta^{3}\Gamma_{int} }
{\delta\Phi'(x^{(3)},\theta_{A}^{(3)})
\delta\Phi'(x^{(2)},\theta_{A}^{(2)})
\delta\Phi'(x^{(1)},\theta_{A}^{(1)})}
\nonumber \\[2pt]
&&= 
\int \prod_{i=1}^{3} dp^{(i)}
\ e^{-i(p^{(1)}x^{(1)}+p^{(2)}x^{(2)}+p^{(3)}x^{(3)})}\
\delta^{2}_{P} (p^{(1)}+p^{(2)}+p^{(3)}) \nonumber \\[2pt]
&& \times \ 
\frac{g}{2}\biggl[
\prod_{A}(\theta^{(1)p^{(3)}}_{A}-\theta^{(2)}_{A})
(\theta^{(1)-p^{(2)}}_{A}-\theta^{(3)}_{A})
+\prod_{A}(\theta^{(1)-p^{(3)}}_{A}-\theta^{(2)}_{A})
(\theta^{(1)p^{(2)}}_{A}-\theta^{(3)}_{A})
\biggr]. \qquad \label{vertex_phi}
\end{eqnarray}
Likewise for $\Gamma_{\oPhi\oPhi\oPhi}$
we obtain,
\begin{eqnarray}
&&\hspace{-20pt}\Gamma_{\oPhi\oPhi\oPhi}^{(tree)} 
(x^{(1)},\theta_{A}^{(1)};x^{(2)},\theta_{A}^{(2)};x^{(3)},\theta_{A}^{(3)})
\nonumber \\[2pt]
&&=\frac{\delta^{3}\Gamma_{int}}
{\delta\oPhi'(x^{(3)},\otheta_{A}^{(3)})
\delta\oPhi'(x^{(2)},\otheta_{A}^{(2)})
\delta\oPhi'(x^{(1)},\otheta_{A}^{(1)})}
\nonumber \\[2pt]
&&= 
\int \prod_{i=1}^{3} dp^{(i)}
\ e^{-i(p^{(1)}x^{(1)}+p^{(2)}x^{(2)}+p^{(3)}x^{(3)})}\
\delta^{2}_{P} (p^{(1)}+p^{(2)}+p^{(3)}) \nonumber \\[2pt]
&& \times \ 
\frac{g}{2}\biggl[
\prod_{A}(\otheta^{(1)p^{(3)}}_{A}-\otheta^{(2)}_{A})
(\otheta^{(1)-p^{(2)}}_{A}-\otheta^{(3)}_{A})
+\prod_{A}(\otheta^{(1)-p^{(3)}}_{A}-\otheta^{(2)}_{A})
(\otheta^{(1)p^{(2)}}_{A}-\otheta^{(3)}_{A})
\biggr], \qquad \label{vertex_ophi}
\end{eqnarray}
where we 
introduced the notations for the ``dressed" $\otheta_{A}$'s
defined by, for instance,
$\otheta_{A}^{(1)p^{(3)}} \equiv 
\otheta^{(1)}_{A} e^{i\oa_{A}p^{(3)}} 
=\otheta^{(1)}_{A}e^{i(\oa_{A})_{\mu}p^{(3)}_{\mu}}$
($A$ : no sum)
with $\oa_{A}=(a,\ta)$.
In deriving these expressions, it is convenient to use
the cyclic properties of the superfields (\ref{int1})-(\ref{int2}).
It is important to remark that in the continuum limit, 
the phase factors in the ``dressed"
$\theta_{A}$'s and $\otheta_{A}$'s
become unity,
so that the factors in the square bracket in
the vertex functions (\ref{vertex_phi})
and (\ref{vertex_ophi}) are reduced into
the ordinary delta functions,
$2\delta^{2}(\theta^{(1)}-\theta^{(2)})
\delta^{2}(\theta^{(1)}-\theta^{(3)})$
and
$2\delta^{2}(\otheta^{(1)}-\otheta^{(2)})
\delta^{2}(\otheta^{(1)}-\otheta^{(3)})$,
respectively.
On the other hand,
on the lattice, the first and the second
term in each square bracket 
give rise to the opposite phases each other
which essentially stems from the non-commutativity between the superfields,
$\Phi'(1)*\Phi'(2)\neq \Phi'(2)*\Phi'(1)$, etc..
This feature is implying a crucial distinction between 
the coherent and non-coherent multiplications
of the phase factors in the perturbative calculations.
In the next section, 
by investigating the matrix valued Wess-Zumino model,
we will explicitly see 
that it corresponds to the
distinction between the planar and non-planar diagrams.

Before finishing this subsection, we note again that
in the above expressions, 
whenever the physical external 
lines are concerned,
the momentum $p^{(i)}$ should be understood as 
the physical momentum plus corresponding momentum region 
$p^{(i)}=k^{(i)}+2\pi_{K^{(i)}}+\pi_{L^{(i)}}$
as defined in (\ref{physical_momentum}). 
The momentum integration should  accordingly be regarded as
the physical momentum integration summed up over all the divisions 
$\pi_{K}$ and $\pi_{L^{(i)}}$,
$\int dp^{(i)} = \sum_{K^{(i)},L^{(i)}}\int_{-\frac{\pi}{2}}^{+\frac{\pi}{2}}
 \frac{d^{2}k^{(i)}}{(4\pi)^{2}}$.
In contrast, 
as far as the internal lines are concerned,
since $k^{(i)}$, $\pi_{K^{(i)}}$ and 
$\pi_{L^{(i)}}$ always come up together
and we eventually sum up all the momentum divisions $\pi_{K^{(i)}}$
and $\pi_{L^{(i)}}$
for each internal momentum $k^{(i)}$,
we are practically allowed to use the notation $p^{(i)}$
and to integrate over the entire momentum region from $-2\pi$ to $2\pi$,
keeping in mind that all the 16 physical momenta are actually contributing
to each internal line.


\section{$\cN=D=2$ Wess-Zumino model with global $U(N)$}

\indent

\begin{figure}

\hspace{20pt}
\includegraphics[width=40mm]{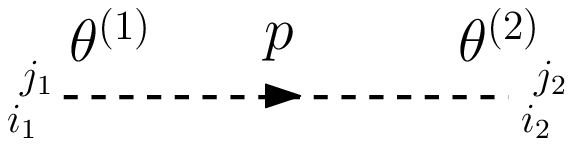}
\hspace{40pt}
$\displaystyle 
\delta^{j_{2}}_{i_{1}} \delta^{j_{1}}_{i_{2}}\, 
\frac{-m}{\sin^{2}p_{\mu}+m^{2}}\delta^{2}(\theta^{(1)}-\theta^{(2)})$  

\hspace{20pt}
\includegraphics[width=40mm]{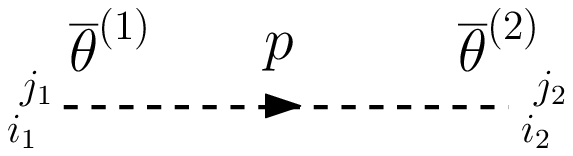}
\hspace{40pt}
$\displaystyle 
\delta^{j_{2}}_{i_{1}} \delta^{j_{1}}_{i_{2}}\, 
\frac{-m}{\sin^{2}p_{\mu}+m^{2}}\delta^{2}(\otheta^{(1)}-\otheta^{(2)})$

\hspace{20pt}
\includegraphics[width=40mm]{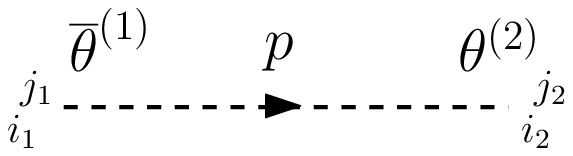}
\hspace{40pt}
$\displaystyle 
\delta^{j_{2}}_{i_{1}} \delta^{j_{1}}_{i_{2}}\, 
\frac{e^{-E^{(12)}_{\mu}\sin p_{\mu}}}
{\sin^{2}p_{\mu}+m^{2}}
\ \ \ 
\textrm{with} \ \ \  
E^{(12)}_{\mu} = \theta^{(1)}\theta^{(2)}_{\mu}
+\epsilon_{\mu\nu}\ttheta^{(1)}\theta^{(2)}_{\nu} $

\hspace{20pt}
\includegraphics[width=40mm]{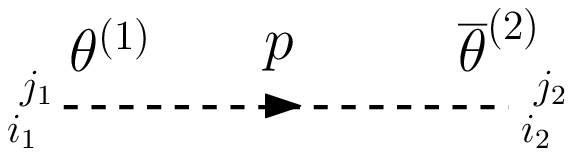}
\hspace{40pt}
$\displaystyle 
\delta^{j_{2}}_{i_{1}} \delta^{j_{1}}_{i_{2}}\, 
\frac{e^{+E^{(21)}_{\mu}\sin p_{\mu}}}
{\sin^{2}p_{\mu}+m^{2}} 
\ \ \ 
\textrm{with} \ \ \  
E^{(21)}_{\mu} = \theta^{(2)}\theta^{(1)}_{\mu}
+\epsilon_{\mu\nu}\ttheta^{(2)}\theta^{(1)}_{\nu}$
\caption{Feynman rules for superfield propagators 
for the $U(N)$ Wess-Zumino model in the matrix notation (\ref{matrix_notation}) :
the momentum $p_{\mu}$ should be understood as
$(k+2\pi_{K}+\pi_{L})_{\mu}$ if external lines are concerned.}
\label{N_fig_propagators}
\end{figure}

\begin{figure}
\hspace{30pt}
\includegraphics[width=35mm]{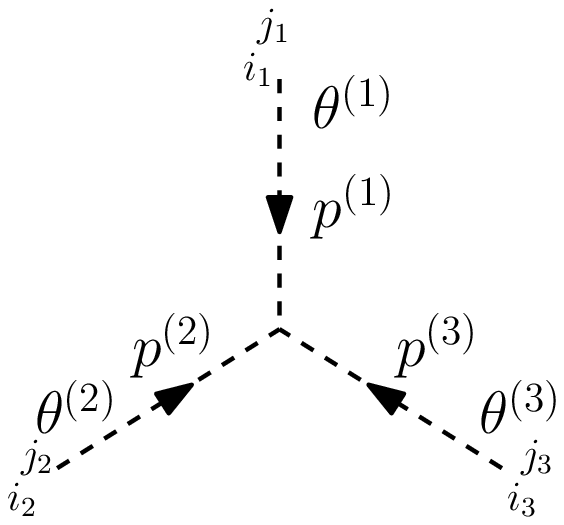}

\vspace{-80pt}
\hspace{178pt}
\parbox{100mm}{
$\displaystyle  
\delta^{2}(p^{(1)}+p^{(2)}+p^{(3)}) \\[2pt]
\times \frac{g}{2}
\biggl[ \delta^{j_{3}}_{i_{1}}\delta^{j_{1}}_{i_{2}}\delta^{j_{2}}_{i_{3}}
\prod_{A}(\theta^{(1)p^{(3)}}_{A}-\theta^{(2)}_{A})
(\theta^{(1)-p^{(2)}}_{A}-\theta^{(3)}_{A}) \\
\hspace{15pt}
+ \delta^{j_{2}}_{i_{1}}\delta^{j_{1}}_{i_{3}}\delta^{j_{3}}_{i_{2}}
\prod_{A}(\theta^{(1)-p^{(3)}}_{A}-\theta^{(2)}_{A})
(\theta^{(1)p^{(2)}}_{A}-\theta^{(3)}_{A}) 
\biggr]
$}

\hspace{30pt}
\includegraphics[width=35mm]{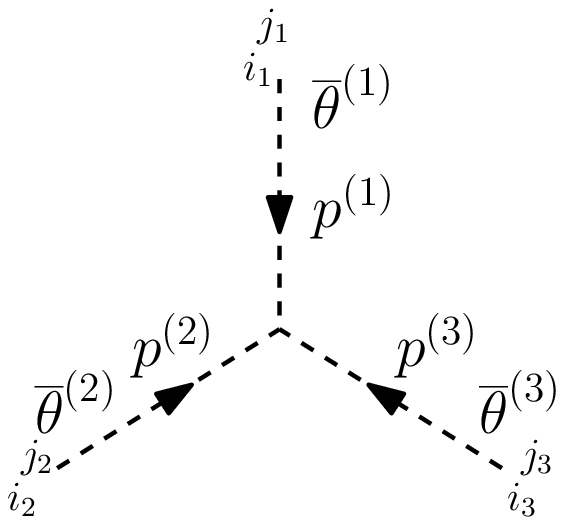}

\vspace{-80pt}
\hspace{178pt}
\parbox{100mm}{
$\displaystyle  
\delta^{2}(p^{(1)}+p^{(2)}+p^{(3)}) \\[2pt]
\times \frac{g}{2}
\biggl[ \delta^{j_{3}}_{i_{1}}\delta^{j_{1}}_{i_{2}}\delta^{j_{2}}_{i_{3}}
\prod_{A}(\otheta^{(1)p^{(3)}}_{A}-\otheta^{(2)}_{A})
(\otheta^{(1)-p^{(2)}}_{A}-\otheta^{(3)}_{A}) \\
\hspace{15pt}
+ \delta^{j_{2}}_{i_{1}}\delta^{j_{1}}_{i_{3}}\delta^{j_{3}}_{i_{2}}
\prod_{A}(\otheta^{(1)-p^{(3)}}_{A}-\otheta^{(2)}_{A})
(\otheta^{(1)p^{(2)}}_{A}-\otheta^{(3)}_{A}) 
\biggr]
$}

\vspace{10pt}
\caption{Feynman rules for the superfield vertices 
for the $U(N)$ Wess-Zumino model in the matrix notation (\ref{matrix_notation}) :
the momentum $p^{(i)}_{\mu}$ should be understood as
$(k+2\pi_{K^{(i)}}+\pi_{L^{(i)}})_{\mu}$ if external lines are concerned.}
\label{N_fig_vertex}
\end{figure}

Since we have derived the propagators and vertex functions, 
it is now straightforward to investigate
the possible radiative corrections for the lattice Wess-Zumino model
(\ref{total_action}).
However, as we have mentioned in the last subsection,
the non-commutativity of the superfields may come up 
with the crucial distinction between 
the planar and non-planar diagrams,
corresponding to the coherent and non-coherent 
multiplications of the phase factors, respectively.
In order to look at this aspect more carefully, it is appropriate
to consider a possible matrix extension of the model (\ref{total_action}).
In this section, we introduce a $N\times N$ matrix version of 
$\cN=D=2$ twisted Wess-Zumino model
which possesses $U(N)$ global symmetry.
We explicitly study the one-loop (in some cases any loops)
quantum corrections to the lattice action.
The model of our interest is given by
\begin{eqnarray}
S^{N\times N}_{WZ} &=&
S^{N\times N}_{kin}
+S^{N\times N}_{mass}
+S^{N\times N}_{int}, \label{N_action}
\end{eqnarray}
with
\begin{eqnarray}
S^{N\times N}_{kin} &=& 
\sum_{x} \int d^{4}\theta
\ \mathrm{Tr} \ 
\oPhi * \Phi, \\[2pt]
S^{N\times N}_{mass} &=& 
\frac{m}{2}
\sum_{x} \biggl[
\int d^{2}\theta
\ \mathrm{Tr} \ 
\Phi * \Phi + 
\int d^{2}\otheta
\ \mathrm{Tr} \ 
\oPhi * \oPhi 
\biggr],
\\[2pt]
S^{N\times N}_{int} &=& 
\frac{g}{3!}
\sum_{x} \biggl[
\int d^{2}\theta
\ \mathrm{Tr} \ 
\Phi * \Phi * \Phi + 
\int d^{2}\otheta
\ \mathrm{Tr} \ 
\oPhi * \oPhi * \oPhi
\biggr],
\end{eqnarray}
where the matrix (anti-)chiral superfields are defined by 
$\Phi (x,\theta,\otheta)= \Phi^{a}(x,\theta,\otheta)T^{a}$ and
$\oPhi (x,\theta,\otheta)= \oPhi^{a}(x,\theta,\otheta)T^{a}$
with the hermitian generators $T^{a}\ (a=1,\dots, N^{2})$ of $U(N)$. 
The action is obviously invariant under
the $U(N)$ global transformation,
\begin{eqnarray}
\Phi(x,\theta,\otheta) \rightarrow \Omega \Phi(x,\theta,\otheta) \Omega^{-1},
\hspace{20pt}
\oPhi(x,\theta,\otheta) \rightarrow \Omega \oPhi(x,\theta,\otheta) \Omega^{-1},
\ \ \mathrm{with} \ \
\Omega \in U(N).
\end{eqnarray} 
Since all the terms are expressed in terms of the star product,
the action is also manifestly invariant under the $\cN=D=2$ twisted SUSY
transformations,
\begin{eqnarray}
\delta_{A} \Phi^{a} &=& [\xi_{A}Q_{A},\Phi^{a}]_{*},
\end{eqnarray} 
where the supercharges $Q_{A}$'s are given by (\ref{Q1})-(\ref{Q3}).
Deriving the component-wise SUSY transformations 
for the matrix superfields is just in a straightforward manner.
Notice that the cyclic permutation property of the superfields
under the trace
is compatible with the cyclic nature of the star product 
listed in (\ref{Dterm})-(\ref{int2}).
One can follow exactly the same procedures as given in the last section,
and derive the corresponding Feynman rules 
for the matrix action (\ref{N_action}).
If one employs the matrix index notation of the superfields,
\begin{eqnarray}
\Phi^{\ j}_{i} &=& \Phi^{a}(T^{a})^{\ j}_{i}, \hspace{20pt}
\oPhi^{\ j}_{i} \ =\ \oPhi^{a}(T^{a})^{\ j}_{i}, 
\label{matrix_notation}
\end{eqnarray}
the $U(N)$ matrix extension of 
the functional derivative relations 
(\ref{phi_derivative}) are expressed as
\begin{eqnarray}
\frac{\delta \Phi'^{\ j_{2}}_{i_{2}}(2)}{\delta \Phi'^{\ j_{1}}_{i_{1}}(1)} 
&=& \delta^{i_{1}}_{i_{2}}\delta^{j_{2}}_{j_{1}}\,
\delta^{2}_{x^{(1)},x^{(2)}}\delta^{2}(\theta^{(1)}-\theta^{(2)}),
\hspace{10pt}
\frac{\delta \oPhi'^{\ j_{2}}_{i_{2}}(2)}{\delta \oPhi'^{\ j_{1}}_{i_{1}}(1)} 
\ =\ \delta^{i_{1}}_{i_{2}}\delta^{j_{2}}_{j_{1}}\,
\delta^{2}_{x^{(1)},x^{(2)}}\delta^{2}(\otheta^{(1)}-\otheta^{(2)}).
\qquad 
\end{eqnarray}
The superfield propagators and the tree level vertex functions
for the $\cN=D=2$ Wess-Zumino model with global $U(N)$ are 
summarized in Fig. \ref{N_fig_propagators} and \ref{N_fig_vertex}.

\begin{figure}
\begin{center}
\includegraphics[width=65mm]{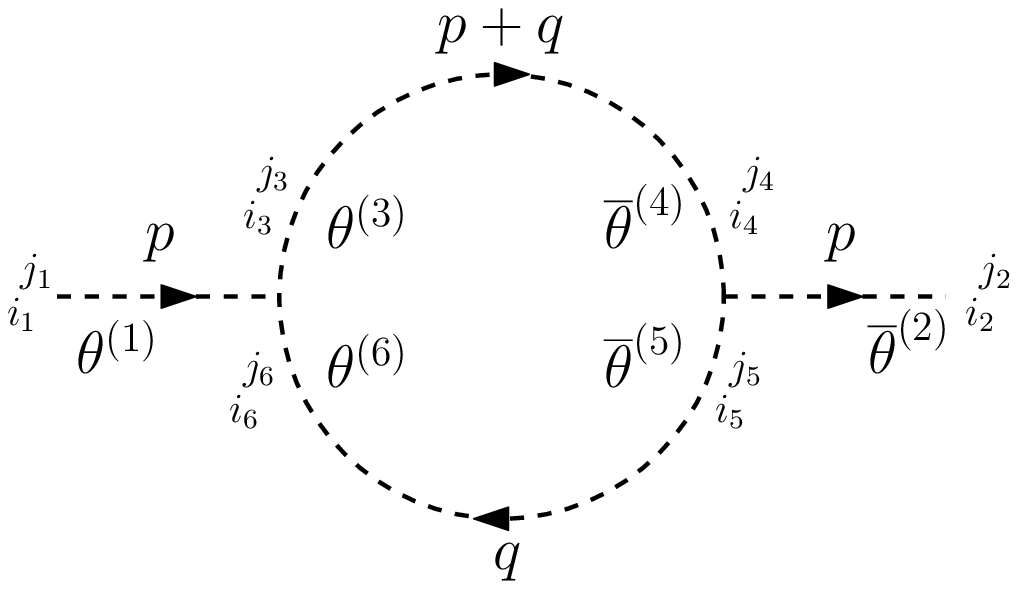}
\caption{One-loop correction to the vertex function $\Gamma_{\Phi\oPhi}$}
\label{fig_phi_ophi}
\end{center}
\end{figure}

Let us first calculate the one-loop self-energy diagram 
shown in Fig. \ref{fig_phi_ophi} which we denote 
$\Gamma_{\Phi\oPhi}^{(1-loop)}$. Up to some overall numerical factors,
it is given by
\begin{eqnarray}
\Gamma_{\Phi\oPhi}^{(1-loop)}
&\sim& 
g^{2} 
\sum_{i_{3}\sim i_{6}} \sum_{j_{3}\sim j_{6}}
\int dq
\int d^{2}\theta^{(3)} d^{2}\otheta^{(4)}
d^{2}\otheta^{(5)} d^{2}\theta^{(6)} 
\nonumber \\
&&\times \ \frac{1}{2}  \biggl[
\delta^{j_{3}}_{i_{1}}\, \delta^{j_{1}}_{i_{6}}\, \delta^{j_{6}}_{i_{3}}\,
\prod_{A}(\theta_{A}^{(1)-(p+q)}-\theta_{A}^{(6)})
(\theta_{A}^{(1)-q}-\theta_{A}^{(3)}) \nonumber \\
&& \hspace{15pt}
+\ \delta^{j_{6}}_{i_{1}}\, \delta^{j_{1}}_{i_{3}}\, \delta^{j_{3}}_{i_{6}}\,
\prod_{A}(\theta_{A}^{(1) p+q}-\theta_{A}^{(6)})
(\theta_{A}^{(1) q}-\theta_{A}^{(3)}) 
\biggr] \nonumber \\
&&\times \ \frac{1}{2} \biggl[
\delta^{j_{5}}_{i_{2}}\, \delta^{j_{2}}_{i_{4}}\, \delta^{j_{4}}_{i_{5}}\,
\prod_{B}(\otheta_{B}^{(2)-q}-\otheta_{B}^{(4)})
(\otheta_{B}^{(2)-(p+q)}-\otheta_{B}^{(5)}) \nonumber \\
&& \hspace{15pt}
+\ \delta^{j_{4}}_{i_{2}}\, \delta^{j_{2}}_{i_{5}}\, \delta^{j_{5}}_{i_{4}}\,
\prod_{B}(\otheta_{B}^{(2) q}-\otheta_{B}^{(4)})
(\otheta_{B}^{(2) p+q}-\otheta_{B}^{(5)}) 
\biggr] \nonumber \\
&&\times\
\delta^{j_{4}}_{i_{3}}\, \delta^{j_{3}}_{i_{4}}\,
\frac{e^{+E^{(43)}_{\mu}\sin (p+q)_{\mu}}}{\sin^{2}(p+q)_{\mu}+m^{2}}
\ \delta^{j_{6}}_{i_{5}}\, \delta^{j_{5}}_{i_{6}}\,
\frac{e^{-E^{(56)}_{\mu}\sin q_{\mu}}}{\sin^{2}q_{\mu}+m^{2}}
\end{eqnarray}
where again $E^{(ij)}_{\mu}=\theta^{(i)}\theta^{(j)}_{\mu}
+ \epsilon_{\mu\nu}\ttheta^{(i)}\theta^{(j)}_{\nu}$
and $\int dq = \int^{2\pi}_{-2\pi}\frac{d^{2}q}{(4\pi)^{2}}$.
\begin{figure}
\begin{center}
\begin{minipage}{80mm}
\begin{center}
\includegraphics[width=55mm]{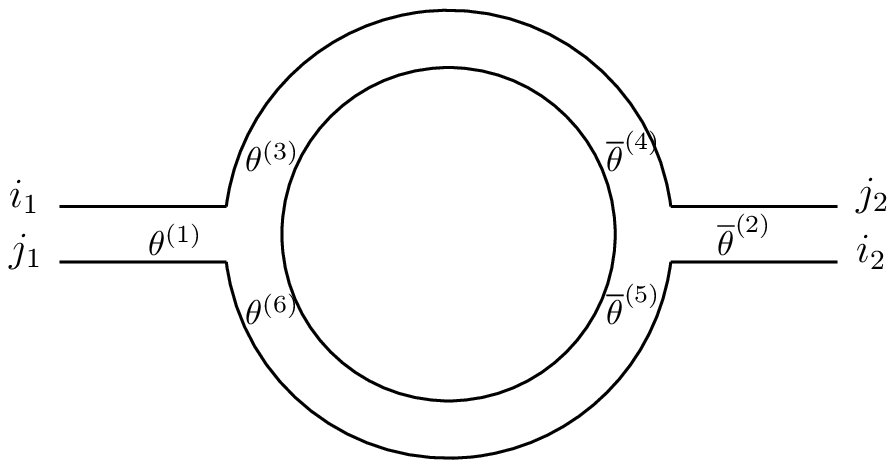}
\caption{Planar contribution to $\Gamma_{\Phi\oPhi}^{(1-loop)}$}
\label{fig_phi_ophi_planar}
\end{center}
\end{minipage}
\begin{minipage}{80mm}
\begin{center}
\includegraphics[width=55mm]{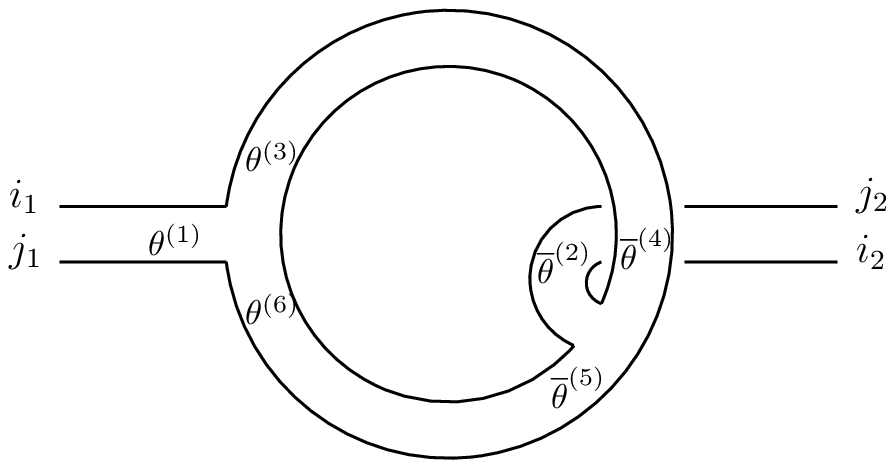}
\caption{Non-planar contribution to $\Gamma_{\Phi\oPhi}^{(1-loop)}$}
\label{fig_phi_ophi_non-planar}
\end{center}
\end{minipage}
\end{center}
\end{figure}
After contracting the matrix indices, one finds that 
the one-loop self-energy
$\Gamma_{\Phi\oPhi}^{(1-loop)}$ can be divided into two parts:
the planar part (Fig. \ref{fig_phi_ophi_planar}) 
and the non-planar part (Fig. \ref{fig_phi_ophi_non-planar}),
\begin{eqnarray}
\Gamma_{\Phi\oPhi}^{(1-loop)} &=&
\Gamma_{\Phi\oPhi(planar)}^{(1-loop)}
+\Gamma_{\Phi\oPhi(non-planar)}^{(1-loop)},
\end{eqnarray}
which are respectively given by
\begin{eqnarray}
\Gamma_{\Phi\oPhi(planar)}^{(1-loop)}
&=& \delta^{j_{2}}_{i_{1}}\, \delta^{j_{1}}_{i_{2}}\,
\frac{g^{2}N}{4}
\int dq
\int d^{2}\theta^{(3)} d^{2}\otheta^{(4)}
d^{2}\otheta^{(5)} d^{2}\theta^{(6)}  \nonumber \\
&& \biggl[
\prod_{A}(\theta_{A}^{(1)-(p+q)}-\theta_{A}^{(6)})
(\theta_{A}^{(1)-q}-\theta_{A}^{(3)})
\prod_{B}(\otheta_{B}^{(2)-q}-\otheta_{B}^{(4)})
(\otheta_{B}^{(2)-(p+q)}-\otheta_{B}^{(5)}) \nonumber \\
&&+
\prod_{A}(\theta_{A}^{(1) p+q}-\theta_{A}^{(6)})
(\theta_{A}^{(1) q}-\theta_{A}^{(3)}) 
\prod_{B}(\otheta_{B}^{(2) q}-\otheta_{B}^{(4)})
(\otheta_{B}^{(2) p+q}-\otheta_{B}^{(5)}) 
\biggr] \nonumber \\
&&\times\ 
\frac{e^{+E^{(43)}_{\mu}\sin (p+q)_{\mu}}}{\sin^{2}(p+q)_{\mu}+m^{2}}
\frac{e^{-E^{(56)}_{\mu}\sin q_{\mu}}}{\sin^{2}q_{\mu}+m^{2}},
\label{phi_ophi_planar}
\end{eqnarray}
\begin{eqnarray}
\Gamma_{\Phi\oPhi(non-planar)}^{(1-loop)}
&=& \delta^{j_{1}}_{i_{1}}\, \delta^{j_{2}}_{i_{2}}\,
\frac{g^{2}}{4}
\int dq
\int d^{2}\theta^{(3)} d^{2}\otheta^{(4)}
d^{2}\otheta^{(5)} d^{2}\theta^{(6)}  \nonumber \\
&& \biggl[
\prod_{A}(\theta_{A}^{(1)-(p+q)}-\theta_{A}^{(6)})
(\theta_{A}^{(1)-q}-\theta_{A}^{(3)}) 
\prod_{B}(\otheta_{B}^{(2) q}-\otheta_{B}^{(4)})
(\otheta_{B}^{(2) p+q}-\otheta_{B}^{(5)}) \nonumber \\
&& +
\prod_{A}(\theta_{A}^{(1) p+q}-\theta_{A}^{(6)})
(\theta_{A}^{(1) q}-\theta_{A}^{(3)}) 
\prod_{B}(\otheta_{B}^{(2)-q}-\otheta_{B}^{(4)})
(\otheta_{B}^{(2)-(p+q)}-\otheta_{B}^{(5)}) 
\biggr] \nonumber \\
&&\times\ 
\frac{e^{+E^{(43)}_{\mu}\sin (p+q)_{\mu}}}{\sin^{2}(p+q)_{\mu}+m^{2}}
\frac{e^{-E^{(56)}_{\mu}\sin q_{\mu}}}{\sin^{2}q_{\mu}+m^{2}}.
\label{phi_ophi_non-planar}
\end{eqnarray}
Note that the planar part has a factor $N$ which stems from
the trace of the matrix indices.
One can define the 't Hooft large-$N$ limit \cite{tHooft}
by rescaling the coupling constant $g^{2}$, namely,
by taking $N\rightarrow \infty$ 
with $\lambda\equiv g^{2}N$ fixed.
In this large-$N$ region, the non-planar contribution (\ref{phi_ophi_non-planar}) 
is suppressed by $\mathcal{O}(\frac{1}{N})$ while
the planar contribution (\ref{phi_ophi_planar})
is order of $\mathcal{O}(1)$.

We shall first evaluate the planar contribution
$\Gamma_{\Phi\oPhi(planar)}^{(1-loop)}$.
After performing the $\theta$ integration,
the first term in the square bracket in (\ref{phi_ophi_planar})
gives rise to the following argument in the exponential,
\begin{eqnarray}
&&[\theta^{(2)-q}\theta^{(1)-q}_{\mu}
+\epsilon_{\mu\nu}\ttheta^{(2)-q}\theta^{(1)-q}_{\nu}]\sin (p+q)_{\mu} 
-[\theta^{(2)-(p+q)}\theta^{(1)-(p+q)}_{\mu}
+\epsilon_{\mu\nu}\ttheta^{(2)-(p+q)}\theta^{(1)-(p+q)}_{\nu}]\sin q_{\mu}
\nonumber \\[2pt]
&&= [\theta^{(2)}\theta^{(1)}_{\mu} e^{-iq_{\mu}}
+\epsilon_{\mu\nu}\ttheta^{(2)}\theta^{(1)}_{\nu}e^{+iq_{\mu}}]\sin (p+q)_{\mu}
-[\theta^{(2)}\theta^{(1)}_{\mu}e^{-i(p+q)_{\mu}}
+\epsilon_{\mu\nu}\ttheta^{(2)}\theta^{(1)}_{\nu}e^{+i(p+q)_{\nu}}]\sin q_{\mu} 
\nonumber \\[2pt]
&&= [\theta^{(2)}\theta^{(1)}_{\mu}+\epsilon_{\mu\nu}\ttheta^{(2)}\theta^{(1)}_{\nu} ]
\sin p_{\mu} \nonumber \\[2pt]
&&= E^{(21)}_{\mu}\sin p_{\mu},
\label{exp_factor1}
\end{eqnarray}
where we used the Leibniz rule conditions
(\ref{N=D=2_Leibniz_cond}) from the first to the second line.
It is straightforward to show that the second term in (\ref{phi_ophi_planar})
also gives rise to the same factor after the Leibniz rule conditions are 
imposed.
The planar contribution to the one-loop self-energy diagram
$\Gamma_{\Phi\oPhi}^{(1-loop)}$
is thus given by,
\begin{eqnarray}
\Gamma_{\Phi\oPhi(planar)}^{(1-loop)}
&=& 
\frac{g^{2}N}{2} 
\delta^{j_{2}}_{i_{1}}\delta^{j_{1}}_{i_{2}}
e^{+E^{(21)}_{\mu}\sin p_{\mu}}
\int^{2\pi}_{-2\pi} \frac{d^{2}q}{(4\pi)^{2}}
\frac{1}{\sin^{2} (p+q)_{\mu}+m^{2}}
\frac{1}{\sin^{2} q_{\mu}+m^{2}}. \qquad
\label{phi_ophi_planar2}
\end{eqnarray}
Since the superspace coordinates are exclusively 
encoded only in the exponential factor $e^{E^{(21)}_{\mu}\sin p_{\mu}}$
in (\ref{phi_ophi_planar2}),
the $\cN=D=2$ twisted SUSY invariant nature of the kinetic part of the action 
(\ref{kinetic_terms})
is strictly protected at non-zero lattice spacing
as long as the planar 
diagram is concerned.
In other words, the supercoordinate structure and
matrix index structure in the superfield propagator
$<\Phi'(1)_{i_{1}}^{\ j_{1}}\oPhi'(2)_{i_{2}}^{\ j_{2}}>$
remains unchanged even after the planar one-loop contribution is
taken into account.
The numerical factor in (\ref{phi_ophi_planar2})
which stems from the $q$-integration may be
absorbed into a wave function renormalization.
These procedures are completely analogous to the ones 
in the continuum theory \cite{Fujikawa-Lang,Wess-Bagger,Piguet-Sibold}.
Notice that these remarkable features are originated from the coherent 
multiplications of the phase factors $e^{ia_{A}p}$ in the planar diagram
as shown in (\ref{exp_factor1}).
In contrast, as one can imagine, the non-planar diagram
$\Gamma_{\Phi\oPhi(non-planar)}^{(1-loop)}$
does not have these favorable features due to 
the non-coherent multiplications of the phase factors.
Actually the first term of (\ref{phi_ophi_non-planar}) would give
rise to the argument of the exponential as,
\begin{eqnarray}
&&[\theta^{(2)q}\theta^{(1)-q}
+\epsilon_{\mu\nu}\ttheta^{(2)q}\theta^{(1)-q}_{\nu}]\sin (p+q)_{\mu} 
-[\theta^{(2)p+q}\theta^{(1)-(p+q)}
+\epsilon_{\mu\nu}\ttheta^{(2)p+q}\theta^{(1)-(p+q)}]\sin q_{\mu}
\nonumber \\[2pt]
&&= E^{(21)}_{\mu}\sin p_{\mu} 
+2ia\cdot(q \sin (p+q)_{\mu}- (p+q)\sin p_{\mu})
(\theta^{(2)}\theta^{(1)}- \epsilon_{\mu\nu}\ttheta^{(2)}\theta^{(1)}_{\nu})
+ \mathcal{O}(a^{2}), \qquad
\end{eqnarray}
while the second term of (\ref{phi_ophi_non-planar}) yields the argument with
the opposite phases,
\begin{eqnarray}
&&[\theta^{(2)-q}\theta^{(1)q}
+\epsilon_{\mu\nu}\ttheta^{(2)-q}\theta^{(1)q}_{\nu}]\sin (p+q)_{\mu} 
-[\theta^{(2)-(p+q)}\theta^{(1)p+q}
+\epsilon_{\mu\nu}\ttheta^{(2)-(p+q)}\theta^{(1)p+q}]\sin q_{\mu}
\nonumber \\[2pt]
&&= E^{(21)}_{\mu}\sin p_{\mu} 
-2ia\cdot(q \sin (p+q)_{\mu}- (p+q)\sin p_{\mu})
(\theta^{(2)}\theta^{(1)}- \epsilon_{\mu\nu}\ttheta^{(2)}\theta^{(1)}_{\nu})
+ \mathcal{O}(a^{2}). \qquad
\end{eqnarray}
Although the $\mathcal{O}(a)$ contributions are canceling each other,
the non-planar diagram would generically 
spoil the SUSY structure at $\mathcal{O}(a^{2})$. 
In the 't Hooft large-$N$ region, however, these SUSY breaking 
contributions in $\Gamma_{\Phi\oPhi(non-planar)}^{(1-loop)}$ are suppressed
by $\mathcal{O}(a^{2}/N)$, since the coupling constant
$g^{2}$ scales as $\frac{1}{N}$.
We thus state that 
at one-loop level,
$\cN=D=2$ twisted SUSY is strictly protected
at non-zero lattice spacing 
in the self-energy vertex  
$\Gamma_{\Phi\oPhi}^{(1-loop)}$
only in the planar limit, or equivalently,
in the 't Hooft large-$N$ limit.

\begin{figure}
\begin{center}
\includegraphics[width=65mm]{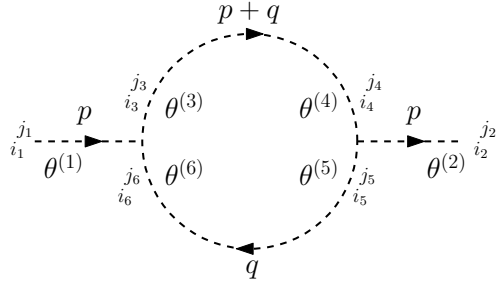}
\caption{One-loop correction to the vertex function $\Gamma_{\Phi\Phi}$}
\label{fig_phi_phi}
\end{center}
\end{figure}

We next evaluate the one-loop correction to the vertex function 
$\Gamma_{\Phi\Phi}$ (Fig. \ref{fig_phi_phi}) which is given by
\begin{eqnarray}
\Gamma_{\Phi\Phi}^{(1-loop)}
&\sim&
g^{2} 
\sum_{i_{3}\sim i_{6}} \sum_{j_{3}\sim j_{6}}
\int dq
\int d^{2}\theta^{(3)} d^{2}\theta^{(4)}
d^{2}\theta^{(5)} d^{2}\theta^{(6)} 
\nonumber \\
&&\times \ \frac{1}{2}  \biggl[
\delta^{j_{3}}_{i_{1}}\, \delta^{j_{1}}_{i_{6}}\, \delta^{j_{6}}_{i_{3}}\,
\prod_{A}(\theta_{A}^{(1)-(p+q)}-\theta_{A}^{(6)})
(\theta_{A}^{(1)-q}-\theta_{A}^{(3)}) \nonumber \\
&& \hspace{15pt}
+\ \delta^{j_{6}}_{i_{1}}\, \delta^{j_{1}}_{i_{3}}\, \delta^{j_{3}}_{i_{6}}\,
\prod_{A}(\theta_{A}^{(1) p+q}-\theta_{A}^{(6)})
(\theta_{A}^{(1) q}-\theta_{A}^{(3)}) 
\biggr] \nonumber \\
&&\times \ \frac{1}{2} \biggl[
\delta^{j_{5}}_{i_{2}}\, \delta^{j_{2}}_{i_{4}}\, \delta^{j_{4}}_{i_{5}}\,
\prod_{B}(\theta_{B}^{(2)-q}-\theta_{B}^{(4)})
(\theta_{B}^{(2)-(p+q)}-\theta_{B}^{(5)}) \nonumber \\
&& \hspace{15pt}
+\ \delta^{j_{4}}_{i_{2}}\, \delta^{j_{2}}_{i_{5}}\, \delta^{j_{5}}_{i_{4}}\,
\prod_{B}(\theta_{B}^{(2) q}-\theta_{B}^{(4)})
(\theta_{B}^{(2) p+q}-\theta_{B}^{(5)}) 
\biggr] \nonumber \\
&&\times\
\delta^{j_{4}}_{i_{3}}\, \delta^{j_{3}}_{i_{4}}\,
\frac{-m\delta^{2}(\theta^{(3)}-\theta^{(4)})}{\sin^{2}(p+q)_{\mu}+m^{2}}
\ \delta^{j_{6}}_{i_{5}}\, \delta^{j_{5}}_{i_{6}}\,
\frac{-m\delta^{2}(\theta^{(5)}-\theta^{(6)})}{\sin^{2}q_{\mu}+m^{2}}.
\qquad
\end{eqnarray}
\begin{figure}
\begin{center}
\begin{minipage}{80mm}
\begin{center}
\includegraphics[width=55mm]{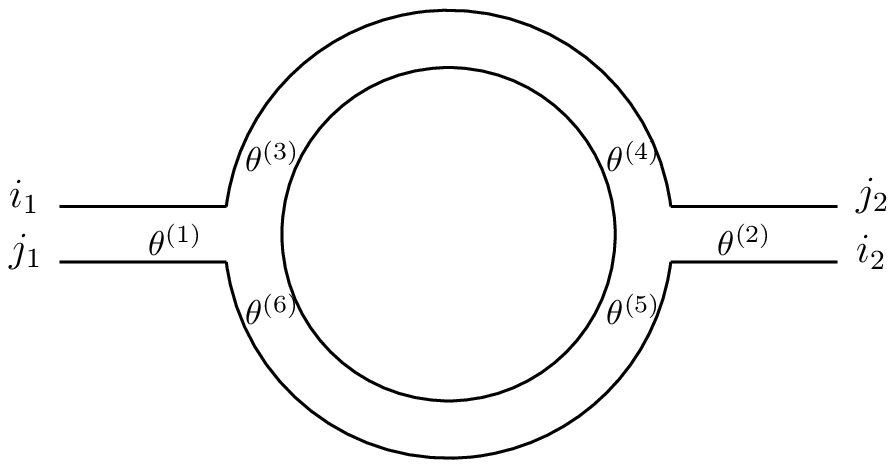}
\caption{Planar contribution to $\Gamma_{\Phi\Phi}^{(1-loop)}$}
\label{fig_phi_phi_planar}
\end{center}
\end{minipage}
\begin{minipage}{80mm}
\begin{center}
\includegraphics[width=55mm]{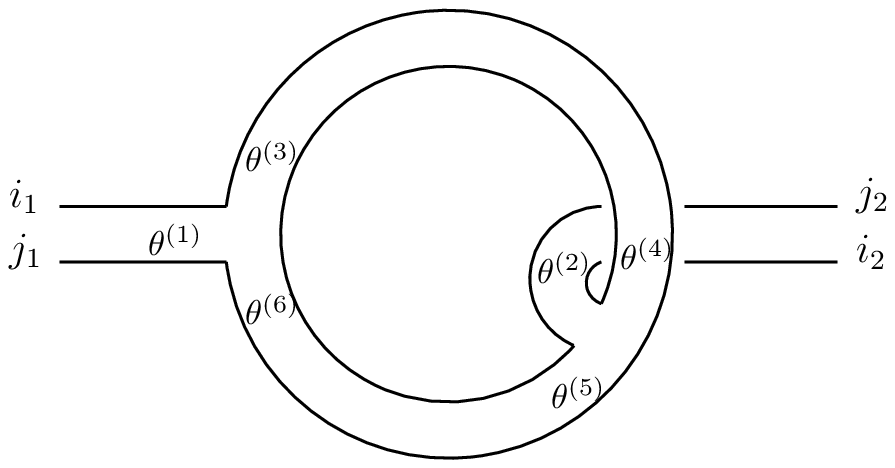}
\caption{Non-planar contribution to $\Gamma_{\Phi\Phi}^{(1-loop)}$}
\label{fig_phi_phi_non-planar}
\end{center}
\end{minipage}
\end{center}
\end{figure}
Following the same procedure as in $\Gamma_{\Phi\oPhi}^{(1-loop)}$,
we can divide the vertex diagram $\Gamma_{\Phi\Phi}^{(1-loop)}$ 
into two parts: 
the planar diagram (Fig. \ref{fig_phi_phi_planar}) 
and the non-planar diagram (Fig. \ref{fig_phi_phi_non-planar}),
\begin{eqnarray}
\Gamma_{\Phi\Phi}^{(1-loop)} &=&
\Gamma_{\Phi\Phi(planar)}^{(1-loop)} 
+ \Gamma_{\Phi\Phi(non-planar)}^{(1-loop)}. 
\end{eqnarray}
After performing the $\theta$ integrations, one finds
that integrand in the planar diagrams
is proportional to,
\begin{eqnarray}
\Gamma_{\Phi\Phi(planar)}^{(1-loop)}
&\propto&
\prod_{A}(\theta_{A}^{(1)-(p+q)}-\theta_{A}^{(2)-(p+q)})
(\theta_{A}^{(1)-q}-\theta_{A}^{(2)-q}) 
+
\prod_{A}(\theta_{A}^{(1)p+q}-\theta_{A}^{(2)p+q})
(\theta_{A}^{(1)q}-\theta_{A}^{(2)q}) \nonumber \\
&=& \prod_{A}(\theta_{A}^{(1)}-\theta_{A}^{(2)})
(\theta_{A}^{(1)}-\theta_{A}^{(2)}) 
+ \prod_{A}(\theta_{A}^{(1)}-\theta_{A}^{(2)})
(\theta_{A}^{(1)}-\theta_{A}^{(2)}) \nonumber \\
& =& 0, \label{phi_phi_planar}
\end{eqnarray}
where we used the relation $a_{1}+a_{2}=0$ and
factored out the overall phases
along the same procedure
as in (\ref{theta_prod_property}).
We thus have the vanishing contribution from the planar diagram.
Note that this planar calculation is completely analogous to the
calculation in the continuum spacetime.
Namely, $\Gamma_{\Phi\Phi}^{(1-loop)}$ should vanish
if the exact SUSY invariance was really maintained. 
In contrast, 
the integrand in the non-planar diagram  
is proportional to, after the $\theta$ integrations, 
\begin{eqnarray}
\Gamma_{\Phi\Phi(non-planar)}^{(1-loop)}
&\propto&\prod_{A}(\theta_{A}^{(1)-(p+q)}-\theta_{A}^{(2)p+q})
(\theta_{A}^{(1)-q}-\theta_{A}^{(2)q})
+ \prod_{A}(\theta_{A}^{(1)p+q}-\theta_{A}^{(2)-(p+q)})
(\theta_{A}^{(1)q}-\theta_{A}^{(2)-q}) \nonumber \\ 
&=& 8\ \theta^{(1)}_{1}\theta^{(2)}_{1} \theta^{(1)}_{2}\theta^{(2)}_{2}
\sin (a_{1}\cdot p)\, 
\sin (a_{2}\cdot p) 
\ \sim\  \mathcal{O}{(a^{2})},
\end{eqnarray}
which implies that 
the non-planar contribution generally spoils 
the exact SUSY structure for the non-zero external momentum $p$.
In the large-$N$ region, however, this contribution 
is again suppressed
by $\mathcal{O}(a^{2}/N)$
as in the case of $\Gamma_{\Phi\oPhi(non-planar)}^{(1-loop)}$.
In a similar way, 
we can show that 
the planar contribution to 
the vertex function of the anti-chiral superfield
$\Gamma_{\oPhi\oPhi}^{(1-loop)}$
manifestly vanish with the same mechanism as the above,
while the non-planar diagram generically spoils the SUSY
structure at $\mathcal{O}(a^{2}/N)$.

\begin{figure}
\begin{center}
\includegraphics[width=65mm]{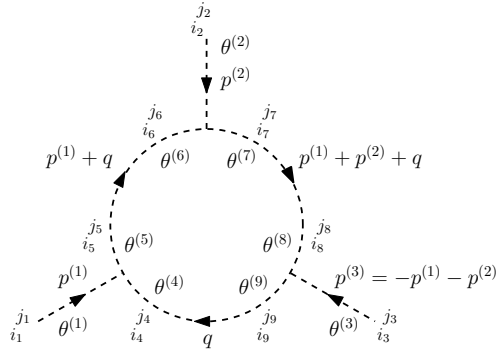}
\caption{One-loop correction to the vertex function $\Gamma_{\Phi\Phi\Phi}$}
\label{fig_phi_phi_phi}
\end{center}
\end{figure}

Let us turn to the one-loop correction to the 3-point vertex function 
$\Gamma_{\Phi\Phi\Phi}$ (Fig. \ref{fig_phi_phi_phi})
which is given by
\begin{eqnarray}
\Gamma_{\Phi\Phi\Phi}^{(1-loop)}
&\sim&
g^{3} 
\sum_{i_{4}\sim i_{9}} \sum_{j_{4}\sim j_{9}}
\int dq
\int  d^{2}\theta^{(4)} d^{2}\theta^{(5)} d^{2}\theta^{(6)} 
d^{2}\theta^{(7)} d^{2}\theta^{(8)} d^{2}\theta^{(9)}
\nonumber \\
&&\times \ \frac{1}{2}  \biggl[
\delta^{j_{5}}_{i_{1}}\, \delta^{j_{1}}_{i_{4}}\, \delta^{j_{4}}_{i_{5}}\,
\prod_{A}(\theta_{A}^{(1)-(p^{(1)}+q)}-\theta_{A}^{(4)})
(\theta_{A}^{(1)-q}-\theta_{A}^{(5)}) \nonumber \\
&& \hspace{15pt}
+\ \delta^{j_{4}}_{i_{1}}\, \delta^{j_{1}}_{i_{5}}\, \delta^{j_{5}}_{i_{4}}\,
\prod_{A}(\theta_{A}^{(1) p^{(1)}+q}-\theta_{A}^{(4)})
(\theta_{A}^{(1) q}-\theta_{A}^{(5)}) 
\biggr] \nonumber \\
&&\times \ \frac{1}{2} \biggl[
\delta^{j_{7}}_{i_{2}}\, \delta^{j_{2}}_{i_{6}}\, \delta^{j_{6}}_{i_{7}}\,
\prod_{B}(\theta_{B}^{(2)-(p^{(1)}+p^{(2)}+q)}-\theta_{B}^{(6)})
(\theta_{B}^{(2)-(p^{(1)}+q)}-\theta_{B}^{(7)}) \nonumber \\
&& \hspace{15pt}
+\ \delta^{j_{6}}_{i_{2}}\, \delta^{j_{2}}_{i_{7}}\, \delta^{j_{7}}_{i_{6}}\,
\prod_{B}(\theta_{B}^{(2) p^{(1)}+p^{(2)}+q}-\theta_{B}^{(6)})
(\theta_{B}^{(2) p^{(1)}+q}-\theta_{B}^{(7)}) 
\biggr] \nonumber \\
&&\times \ \frac{1}{2} \biggl[
\delta^{j_{9}}_{i_{3}}\, \delta^{j_{3}}_{i_{8}}\, \delta^{j_{8}}_{i_{9}}\,
\prod_{C}(\theta_{C}^{(3)-q}-\theta_{C}^{(8)})
(\theta_{C}^{(3)-(p^{(1)}+p^{(2)}+q)}-\theta_{C}^{(9)}) \nonumber \\
&& \hspace{15pt}
+\ \delta^{j_{8}}_{i_{3}}\, \delta^{j_{3}}_{i_{9}}\, \delta^{j_{9}}_{i_{8}}\,
\prod_{C}(\theta_{C}^{(3) q}-\theta_{C}^{(8)})
(\theta_{C}^{(3) p^{(1)} + p^{(2)}+q}-\theta_{C}^{(9)}) 
\biggr] \nonumber \\
&&\times\
\delta^{j_{6}}_{i_{5}}\, \delta^{j_{5}}_{i_{6}}\,
\frac{-m\delta^{2}(\theta^{(5)}-\theta^{(6)})}{\sin^{2}(p^{(1)}+q)_{\mu}+m^{2}}
\ \delta^{j_{8}}_{i_{7}}\, \delta^{j_{7}}_{i_{8}}\,
\frac{-m\delta^{2}(\theta^{(7)}-\theta^{(8)})}
{\sin^{2}(p^{(1)}+p^{(2)}+q)_{\mu}+m^{2}} \nonumber \\
&&\times\ \delta^{j_{4}}_{i_{9}}\, \delta^{j_{9}}_{i_{4}}\,
\frac{-m\delta^{2}(\theta^{(9)}-\theta^{(4)})}
{\sin^{2}q_{\mu}+m^{2}}  \\[4pt]
&=& \Gamma_{\Phi\Phi\Phi(planar)}^{(1-loop)}
+ \Gamma_{\Phi\Phi\Phi(non-planar)}^{(1-loop)}.
\label{gamma_phi_phi_phi}
\end{eqnarray}
\begin{figure}
\begin{center}
\begin{minipage}{80mm}
\begin{center}
\includegraphics[width=50mm]{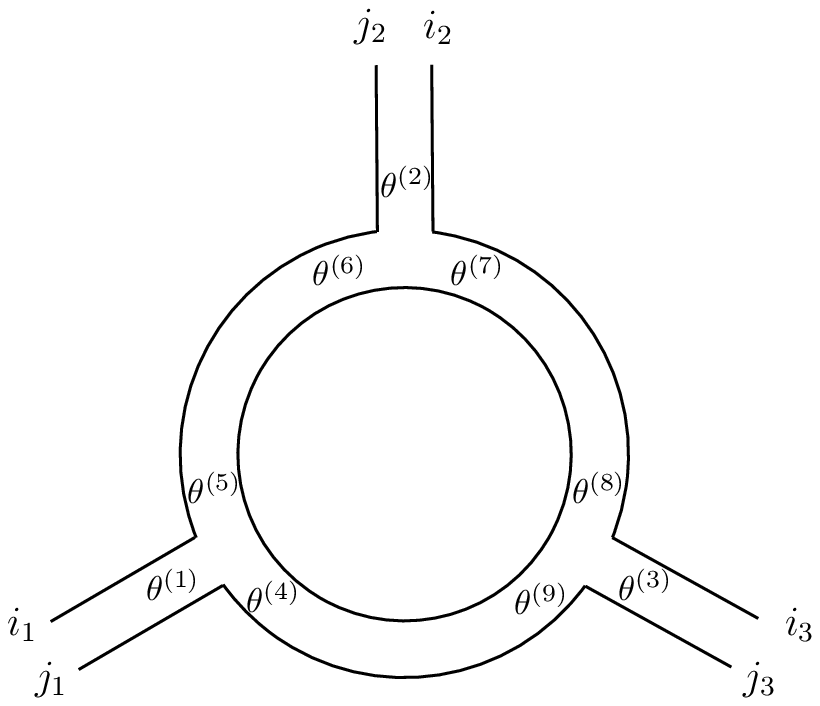}
\caption{Planar contribution to $\Gamma_{\Phi\Phi\Phi}^{(1-loop)}$}
\label{fig_phi_phi_phi_planar}
\end{center}
\end{minipage}
\begin{minipage}{80mm}
\begin{center}
\includegraphics[width=50mm]{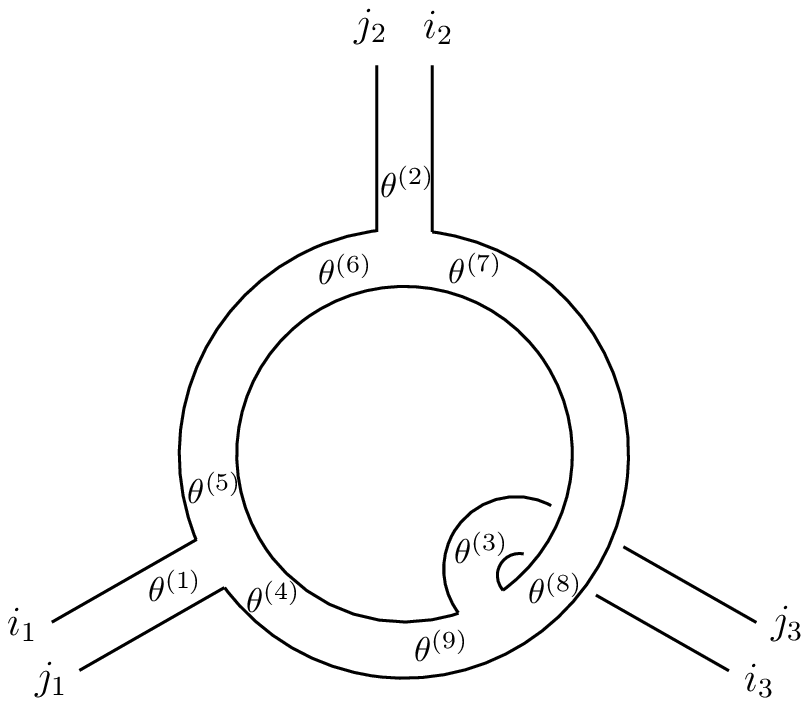}
\caption{Non-planar contribution to $\Gamma_{\Phi\Phi\Phi}^{(1-loop)}$:
we also have the diagrams with 
cyclic permutation of $(i_{1},j_{1}),(i_{2},j_{2}),(i_{3},j_{3})$.} 
\label{fig_phi_phi_phi_non-planar}
\end{center}
\end{minipage}
\end{center}
\end{figure}
Following the same procedure as before,
one can see that the planar contribution 
$\Gamma_{\Phi\Phi\Phi(planar)}^{1-loop}$
(Fig. \ref{fig_phi_phi_phi_planar}), 
which stems from the coherent multiplications of the phase factors
turns out to be vanishing. 
Actually, after the $\theta^{(i)}(i=4\sim 9)$ are integrated, 
it is proportional to,
\begin{eqnarray}
\Gamma_{\Phi\Phi\Phi(planar)}^{(1-loop)}
&\propto&
\prod_{A}(\theta_{A}^{(1)-(p^{(1)}+q)}
-\theta_{A}^{(3)-(p^{(1)}+p^{(2)}+q)})
(\theta_{A}^{(1)-q}
-\theta_{A}^{(2)-(p^{(1)}+p^{(2)}+q)})
(\theta_{A}^{(2)-(p^{(1)}+q)}-\theta_{A}^{(3)-q})
\nonumber  \\
&& +\prod_{A}(\theta_{A}^{(1)p^{(1)}+q}
-\theta_{A}^{(3)p^{(1)}+p^{(2)}+q})
(\theta_{A}^{(1)q}
-\theta_{A}^{(2)p^{(1)}+p^{(2)}+q})
(\theta_{A}^{(2)p^{(1)}+q}-\theta_{A}^{(3)q}) \nonumber \\
&=& 
\prod_{A}(\theta_{A}^{(1)-q}
-\theta_{A}^{(3)-(p^{(2)}+q)})
(\theta_{A}^{(1)-q}
-\theta_{A}^{(2)-(p^{(1)}+p^{(2)}+q)})
(\theta_{A}^{(2)-(p^{(1)}+p^{(2)}+q)}-\theta_{A}^{(3)-(p^{(2)}+q)})
\nonumber  \\
&&+ \prod_{A}(\theta_{A}^{(1)q}
-\theta_{A}^{(3)p^{(2)}+q})
(\theta_{A}^{(1)q}
-\theta_{A}^{(2)p^{(1)}+p^{(2)}+q})
(\theta_{A}^{(2)p^{(1)}+p^{(2)}+q}-\theta_{A}^{(3)p^{(2)}+q}) \nonumber \\
&=& 
\prod_{A}(\theta_{A}^{(1)'}
-\theta_{A}^{(3)'})
(\theta_{A}^{(1)'}
-\theta_{A}^{(2)'})
(\theta_{A}^{(2)'}-\theta_{A}^{(3)'})
\nonumber  \\
&&+\prod_{A}(\theta_{A}^{(1)''}
-\theta_{A}^{(3)''})
(\theta_{A}^{(1)''}
-\theta_{A}^{(2)''})
(\theta_{A}^{(2)''}-\theta_{A}^{(3)''})
\nonumber  \\
&=& 0,  
\label{3-point_planar_phase}
\end{eqnarray}
where from the first to the second line,
we made use of the properties of the $\theta_{A}$ factors
such as (\ref{theta_prod_property}),
while from the second to the third line in (\ref{3-point_planar_phase}), 
we absorbed the phase factors and redefined the $\theta_{A}$'s,
\begin{eqnarray}
\theta_{A}^{(1)'} &=& \theta_{A}^{(1)-q}, \hspace{20pt}
\theta_{A}^{(2)'} \ =\ \theta_{A}^{(2)-(p^{(1)}+p^{(2)}+q)}, \hspace{20pt}
\theta_{A}^{(3)'} \ =\ \theta_{A}^{(3)-(p^{(2)}+q)}, \\[2pt] 
\theta_{A}^{(1)''} &=& \theta_{A}^{(1)q}, \hspace{25pt}
\theta_{A}^{(2)''} \ =\ \theta_{A}^{(2)p^{(1)}+p^{(2)}+q}, \hspace{31pt}
\theta_{A}^{(3)''} \ =\ \theta_{A}^{(3)p^{(2)}+q}, 
\end{eqnarray}
in order to emphasize the manifest nilpotency of $\theta^{(i)'}_{A}$'s and 
$\theta^{(i)''}_{A}$'s. 
On the other hand, the non-planar contribution 
$\Gamma_{\Phi\Phi\Phi(non-planar)}^{(1-loop)}$
(Fig. \ref{fig_phi_phi_phi_non-planar})
turns to be proportional to 
\begin{eqnarray}
\Gamma_{\Phi\Phi\Phi(non-planar)}^{(1-loop)}
&\propto&
8\, \theta^{(1)}_{1} \theta^{(2)}_{1} \theta^{(3)}_{1}
\theta^{(1)}_{2} \theta^{(2)}_{2} \theta^{(3)}_{2} 
\bigl[
\sin a_{1}\cdot (p_{1}+p_{2}) \sin a_{2}\cdot (p_{1}+p_{2}) \nonumber \\
&&+\sin a_{1}\cdot p_{1} \sin a_{2}\cdot p_{1}
+\sin a_{1}\cdot p_{2} \sin a_{2}\cdot p_{2}
\bigr]  \nonumber \\
&\sim& \mathcal{O}(a^{2}),
\end{eqnarray}
which is not vanishing for the generic external momenta $p^{(1)}$ and $p^{(2)}$.
In the large-$N$ region where the coupling constant $g$ is proportional to
$\frac{1}{\sqrt{N}}$,
the SUSY breaking non-planar contributions to the tree point vertex function
$\Gamma_{\Phi\Phi\Phi}^{(1-loop)}$ are thus suppressed by 
$\mathcal{O}(a^{2}/N^{\frac{3}{2}})$.
In a similar manner, one can show that the planar diagrams 
contributing to the one-loop three-point vertex function 
$\Gamma_{\oPhi\oPhi\oPhi}^{1-loop}$
manifestly vanish at non-zero lattice spacings due to the 
same mechanism as $\Gamma_{\Phi\Phi\Phi(planar)}^{(1-loop)}$, 
while the non-planar contributions generically spoils the SUSY structure
at the order of $\mathcal{O}(a^{2}/N^{\frac{3}{2}})$
in the large-$N$ region just as in the case of 
$\Gamma_{\Phi\Phi\Phi(non-planar)}^{(1-loop)}$.

Let us summarize the above results for the 
superfield one-loop calculations
at non-zero lattice spacing.
We have seen that the mass $m$ and coupling constant $g$
in the twisted $\cN=D=2$ Wess-Zumino model 
are strictly protected from the radiative corrections
in the planar diagrams
apart from the wave function renormalization
which stems from the self-energy diagram (\ref{phi_ophi_planar2}).
On the other hand, the non-planar diagrams in general give rise to 
the SUSY spoiling contributions although they are suppressed 
at least at the order of
$\mathcal{O}(a^{2}/N)$ in the 't Hooft large-$N$ region.
The breakdown of twisted SUSY in the non-planar diagrams
is originated from the non-coherent multiplication of the 
phase factors, which stems from the 
intrinsic non-commutative nature of the star product,
$\Phi_{1}*\Phi_{2}\neq \Phi_{2}*\Phi_{1}$.   
These features imply that the exact lattice realization of $\cN=D=2$ twisted SUSY 
in terms of the present non-commutative framework
is achieved in the planar limit,
or equivalently, in the 't Hooft large-$N$ limit.
It should be emphasized that,
as we have seen in the above calculations,
the loop calculations in the planar diagrams
are completely analogous to the ones in the continuum superspace
calculations. 
Therefore, we observe that, 
by further developing the lattice superfield techniques
and  following the argument of 
Grisaru, $\mathrm{Ro\check{c}ek}$ and Siegel \cite{GRS},
a Dirac-K\"ahler twisted $\cN=D=2$ analog of 
the non-renormalization
theorem may also be proven directly on the lattice in the large-$N$ limit.
The result will be given elsewhere.

\begin{figure}
\hspace{30pt}
\includegraphics[width=35mm]{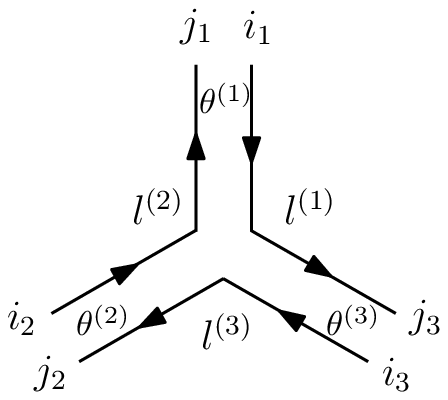}

\vspace{-75pt}
\hspace{148pt}
\parbox{100mm}{
$\displaystyle  
\frac{g}{2}
\biggl[ \delta^{j_{3}}_{i_{1}}\delta^{j_{1}}_{i_{2}}\delta^{j_{2}}_{i_{3}}
\prod_{A}(\theta^{(1)l^{(3)}}_{A}-\theta^{(2)l^{(1)}}_{A})
(\theta^{(1)l^{(3)}}_{A}-\theta^{(3)l^{(2)}}_{A}) \\
\hspace{15pt}
+ \delta^{j_{2}}_{i_{1}}\delta^{j_{1}}_{i_{3}}\delta^{j_{3}}_{i_{2}}
\prod_{A}(\theta^{(1)-l^{(3)}}_{A}-\theta^{(2)-l^{(1)}}_{A})
(\theta^{(1)-l^{(3)}}_{A}-\theta^{(3)-l^{(2)}}_{A}) 
\biggr]
$}
\vspace{20pt}
\caption{Feynman rule for the three-point vertex in the double line notation
with the conserved momenta $l^{(i)}$ }
\label{vertex_double_line}
\end{figure}

\begin{figure}
\begin{center}
(a)
\includegraphics[width=45mm]{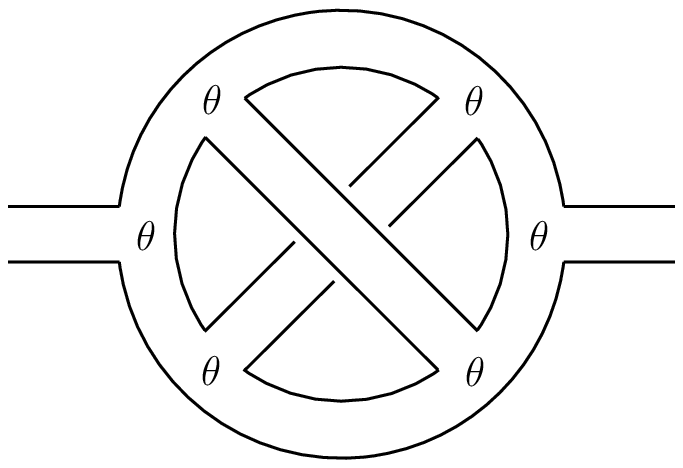}
\hspace{20mm}
(b)
\includegraphics[width=45mm]{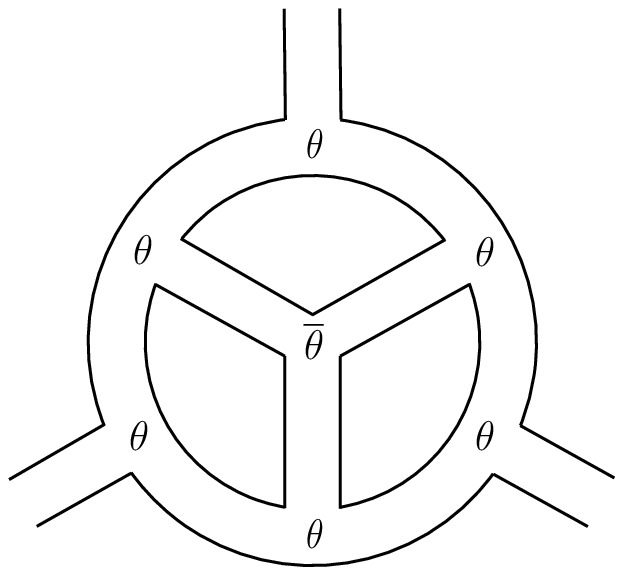}
\end{center}
\caption{Examples of the vanishing diagrams
with (a) non-planarity and (b) propagators $<\Phi\oPhi>$ }
\label{fig_vanishing_example}
\end{figure}

Before finishing this section,
it is instructive to show within the current superfield techniques
that a certain class of planar diagrams
manifestly vanish at any order of perturbation theory
at non-zero lattice spacing.
One can actually see that 
diagrams for $\Gamma_{\Phi\Phi}$ $(\Gamma_{\Phi\Phi})$
and $\Gamma_{\Phi\Phi\Phi}$ $(\Gamma_{\Phi\Phi\Phi})$
manifestly vanish 
if the outer edges of the diagrams 
only contain planar vertices
and the propagators
$<\Phi\Phi>$ $(<\oPhi\oPhi>)$.
In showing such properties, it is convenient to employ the double-line notation
with the conserved momenta $l^{(i)}$'s defined by,
$p^{(1)}=l^{(1)}-l^{(2)}, p^{(2)}=l^{(2)}-l^{(3)}$ and  
$p^{(3)}=l^{(3)}-l^{(1)}$.
Fig. \ref{vertex_double_line} depicts 
the three-point vertex function in this momentum notation. 
\begin{figure}
\begin{center}
\includegraphics[width=80mm]{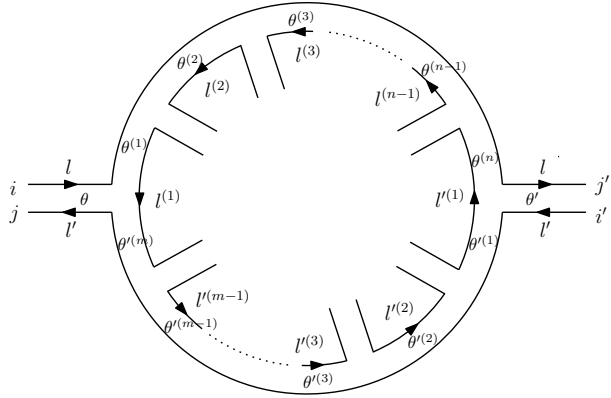}
\caption{Manifestly vanishing diagram for 
the two point vertex function $\Gamma_{\Phi\Phi}$}
\label{fig_phi_phi_planar_all_order}
\end{center}
\end{figure}
Fig. \ref{fig_phi_phi_planar_all_order} depicts the 
class of manifestly vanishing diagrams for the two-point vertex function 
$\Gamma_{\Phi\Phi}$,
which is essentially proportional to,
after performing the $\theta$ integrations
for the propagators, 
\begin{eqnarray}
(\mathrm{Fig.}\ref{fig_phi_phi_planar_all_order})
&\propto&
\int d^{2}\theta^{(1)} d^{2}\theta^{(2)} \cdots d^{2}\theta^{(n)}
\ d^{2}\theta'^{(1)} d^{2}\theta'^{(2)} \cdots d^{2}\theta'^{(m)} 
\nonumber \\[2pt]
&& \times \biggl[
\prod_{A}(\theta_{A}^{l^{(1)}}-\theta_{A}^{(1)l'})
(\theta_{A}^{(1)l^{(2)}}-\theta_{A}^{(2)l^{(1)}})
(\theta_{A}^{(2)l^{(3)}}-\theta_{A}^{(3)l^{(2)}}) 
\nonumber \\[-2pt]
&&\times \cdots \times
(\theta_{A}^{(n-1)l'^{(1)}}-\theta_{A}^{(n)l^{(n-1)}})
(\theta_{A}^{(n)l'}-\theta_{A}'^{l'^{(1)}}) 
\nonumber \\[6pt]
&& \times \prod_{B}(\theta_{B}'^{l'^{(1)}}-\theta_{B}'^{(1)l})
(\theta_{B}'^{(1)l'^{(2)}}-\theta_{B}'^{(2)l'^{(1)}})
(\theta_{B}'^{(2)l'^{(3)}}-\theta_{B}'^{(3)l'^{(2)}}) 
\nonumber \\[-2pt]
&&\times \cdots \times
(\theta_{B}'^{(m-1)l^{(1)}}-\theta_{B}'^{(m)l'^{(m-1)}})
(\theta_{B}'^{(m)l}-\theta_{B}^{l^{(1)}})  
\nonumber \\[4pt]
&&+\ (\mathrm{the\ term\ with\ the\ opposite\ phase} )
\biggr] \nonumber \\[2pt]
&=& 2 \prod_{A}(\theta_{A}-\theta'_{A})
\prod_{B}(\theta_{B}-\theta'_{B})
\nonumber \\[2pt]
&=& 0. \label{2-point_planar_general}
\end{eqnarray}
\begin{figure}
\begin{center}
\includegraphics[width=85mm]{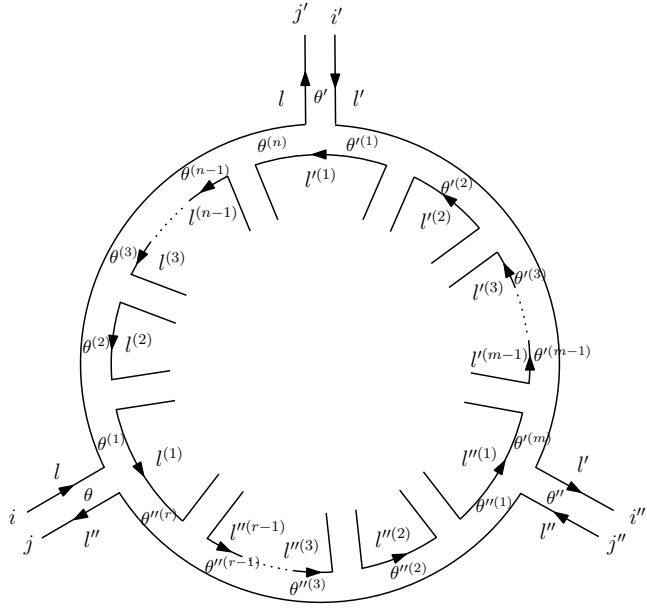}
\caption{Manifestly vanishing diagram for 
the three point vertex function $\Gamma_{\Phi\Phi\Phi}$}
\label{fig_phi_phi_phi_planar_all_order}
\end{center}
\end{figure}
Likewise the diagrams in Fig. \ref{fig_phi_phi_phi_planar_all_order} 
for $\Gamma_{\Phi\Phi\Phi}$ is proportional to,
after integrating the $\theta$'s associated with the propagators,
\begin{eqnarray}
(\mathrm{Fig.}\ref{fig_phi_phi_phi_planar_all_order})
&\propto&
\int d^{2}\theta^{(1)} d^{2}\theta^{(2)} \cdots d^{2}\theta^{(n)}
\ d^{2}\theta'^{(1)} d^{2}\theta'^{(2)} \cdots d^{2}\theta'^{(m)} 
\ d^{2}\theta''^{(1)} d^{2}\theta''^{(2)} \cdots d^{2}\theta''^{(r)} 
\nonumber \\[2pt]
&& \times \biggl[
\prod_{A}(\theta_{A}^{l^{(1)}}-\theta_{A}^{(1)l''})
(\theta_{A}^{(1)l^{(2)}}-\theta_{A}^{(2)l^{(1)}})
(\theta_{A}^{(2)l^{(3)}}-\theta_{A}^{(3)l^{(2)}}) 
\nonumber \\[-2pt]
&&\times \cdots \times
(\theta_{A}^{(n-1)l'^{(1)}}-\theta_{A}^{(n)l^{(n-1)}})
(\theta_{A}^{(n)l'}-\theta_{A}'^{l'^{(1)}}) 
\nonumber \\[6pt]
&& \times \prod_{B}(\theta_{B}'^{l'^{(1)}}-\theta_{B}'^{(1)l})
(\theta_{B}'^{(1)l'^{(2)}}-\theta_{B}'^{(2)l'^{(1)}})
(\theta_{B}'^{(2)l'^{(3)}}-\theta_{B}'^{(3)l'^{(2)}}) 
\nonumber \\[-2pt]
&&\times \cdots \times
(\theta_{B}'^{(m-1)l''^{(1)}}-\theta_{B}'^{(m)l'^{(m-1)}})
(\theta_{B}'^{(m)l''}-\theta_{B}''^{l''^{(1)}})  
\nonumber \\[4pt]
&& \times \prod_{C}(\theta_{C}''^{l''^{(1)}}-\theta_{C}''^{(1)l'})
(\theta_{C}''^{(1)l''^{(2)}}-\theta_{C}''^{(2)l''^{(1)}})
(\theta_{C}''^{(2)l''^{(3)}}-\theta_{C}''^{(3)l''^{(2)}}) 
\nonumber \\[-2pt]
&&\times \cdots \times
(\theta_{C}''^{(r-1)l^{(1)}}-\theta_{C}''^{(r)l''^{(r-1)}})
(\theta_{C}''^{(r)l}-\theta_{C}^{l^{(1)}}) \nonumber \\[2pt]
&&+\ (\mathrm{the\ term\ with\ the\ opposite\ phase} ) 
\biggr]
\nonumber \\[4pt]
&=& 2\prod_{A}(\theta_{A}^{l'}-\theta_{A}'^{l''})
\prod_{B}(\theta_{B}'^{l''}-\theta_{B}''^{l})
\prod_{C}(\theta_{C}''^{l}-\theta_{C}^{l'})
\nonumber \\[2pt]
&=& 0. \label{3-point_planar_general}
\end{eqnarray}
It is worthwhile to
note that the vanishing diagrams 
Fig. \ref{fig_phi_phi_planar_all_order}
and
Fig. \ref{fig_phi_phi_phi_planar_all_order}
generally include  
not only planar diagrams but also a certain class of 
non-planar diagrams
whose non-planarity reside 
inside of the diagrams
such as Fig. \ref{fig_vanishing_example} (a).
Also the diagrams accommodating the propagators $<\Phi\oPhi>$
in their inside such as
Fig. \ref{fig_vanishing_example} (b)
also generically belong to the category of 
manifestly vanishing diagrams.

\section{Summary \& Discussions}

\indent

Having constructed a novel star product formulation of
$\cN=D=2$ twisted SUSY invariant action on a lattice,
we perturbatively studied the possible quantum corrections
at non-zero lattice spacing.
We have explicitly calculated 
the one-loop (in some case any loops) radiative corrections
for $\cN=D=2$ twisted Wess-Zumino model with global $U(N)$ symmetry,
and we have seen that the planar diagrams
strictly protect the entire $\cN=D=2$ twisted
SUSY structure at non-zero lattice spacing.
On the other hand, the non-planar contributions generically
spoil the lattice SUSY although they are suppressed 
at least by the order of $\mathcal{O}(a^{2}/N)$.
From these features we claim that 
the full realization of exact lattice SUSY
is achieved in the 't Hooft large-$N$ limit.


In the series of papers in the collaboration with A. D'Adda, I. Kanamori and 
N. Kawamoto \cite{DKKN1,DKKN2,DKKN3}, 
we have been proposing a theoretical framework to realize exact 
SUSY on a lattice. Since the lattice action
in this framework is given in terms of the superfields, 
their SUSY invariance is formally kept 
manifest at the classical level, 
although its quantum behaviour has not been clarified until this paper. 
This paper addresses how the quantum corrections would spoil or protect 
the classical SUSY 
invariant nature of the lattice action by explicitly calculating the radiative 
corrections to the lattice action perturbatively. As was given in the one-loop 
calculations, (\ref{phi_ophi_planar2}), (\ref{phi_phi_planar}) 
and (\ref{3-point_planar_phase}), the planar diagrams
strictly protect the SUSY 
invariance, although the non-planar diagrams in general spoil the lattice 
SUSY at non-zero lattice spacing. 
This result shows that even though we start from the ``manifest" SUSY 
formulation at classical level, the SUSY invariant nature of the action is generally 
spoiled by the non-planar loop conrtibutions. 
Thus the validity of the manifest lattice 
SUSY formulation could be claimed only in the 't Hooft large-$N$ limit at least 
in the non-commutative product formulation introduced in this paper.
Nevertheless, it is still important to stress that the SUSY protecting nature 
of the planar diagrams is far from accidental 
and is actually completely analogous to the twisted version 
of the superfield calculations in the continuum spacetime
\cite{Fujikawa-Lang}-\cite{Wess-Zumino}. 
This is mainly because 
the non-trivial phase factors associated with each vertex, 
which stem from the
``mild" non-commutativity (\ref{NC2}),
cancel with each other in the planar diagrams.
The generally vanishing planar diagrams 
(\ref{2-point_planar_general}) and 
(\ref{3-point_planar_general}) exhibit this aspect explicitly. 


The consequence of this paper may naturally be understood 
if one reminds the notion of ``proper" ordering appeared
in the following two physically independent contexts.
The first one is, 
as we have mentioned in Sec. 3 and also discussed in Ref. \cite{DKKN3},
that the notion of ``proper" ordering of (super)fields
is essential in the present framework of exact lattice SUSY
due to the non-commutative nature of the 
SUSY transformation on the lattice.
This is originated from the superficial ordering sensitive nature 
of the difference operations on the lattice. 
The second one is, as is well known, that
taking the large-$N$ limit also essentially restricts the ordering of 
the (super)fields in a certain manner,
since interchanging the order of the fields generally
corresponds to converting planar diagrams to 
the corresponding non-planar diagrams or vice versa.
In this paper we have explicitly seen that
respecting the ``proper" ordering in the lattice SUSY context
essentially corresponds to picking up the planar diagrams
in the perturbative expansions.  
In other words, the somewhat peculiar notion of 
``proper" ordering introduced in the context of lattice SUSY realization
now turns to have a physically relevant meaning in the large-$N$ limit.

One might naively think that the 
large-$N$ reduction 
\cite{Eguchi-Kawai}
would make the lattice SUSY problem completely trivial from the beginning
since in the large-$N$ limit
the entire lattice coordinate may be reduced into a single site,
and no chance of the lattice Leibniz rule problem itself may occur.
However, one should notice
that the shift operations or difference operations on the lattice
still remain  encoded in the shifting matrices
even after the reduction,  
for instance,
in $\Gamma_{\mu}$ with
$\Gamma_{\mu}\Gamma_{\nu}=Z_{\nu\mu}\Gamma_{\nu}\Gamma_{\mu}$,
if one takes 
the twisted Eguchi-Kawai reduction \cite{TEK}.  
From this point of view, the lattice SUSY formulation in the large-$N$ limit 
may  naturally be converted into a problem of how to supersymmetrize 
these shift matrices.
We should note that
this is actually the subject which was partially 
addressed in Ref. \cite{ADKS}.

It is interesting to consider if the star product formulation introduced
in this paper could be applied to formulating the supersymmetric 
gauge theories on the lattice. 
We observe that 
the lattice gauge covariant formulation introduced in Ref. \cite{DKKN2,DKKN3}
by means of the link supercharges and the link component fields 
may be extended to a certain type of star gauge covariant superfield formulation
on the lattice.  
Also the lattice Chern-Simons formulation based on the $\cN=4\ D=3$
twisted SUSY which was recently introduced in Ref. \cite{Nagata-Wu}
may be recasted into this category.
It is also worthwhile to 
further develop the lattice superfield framework along the similar manner 
as given in Ref.\cite{GRS}
and try to accomplish the lattice analog of the non-renormalization 
theorem directly at non-zero lattice spacing.

In this paper, all the perturbative calculations are performed
with the symmetric choice of $a_{A}$ (\ref{a_symmetric_choice}).
We have introduced four copies of the lattice superfields
essentially due to this symmetric choice.
This is because the lattice SUSY transformations
involve the mapping of the component fields
from the integer sites to the half-integer sites, or vice versa,
due to the non-commutativity associated with $\theta_{A}$ and $\xi_{A}$.
One may wonder what would happen if one takes any other choices.
In particular, the asymmetric choice where 
the shift parameters are given by, $a=0, a_{\mu}=+n_{\mu}$ and $\ta=-n_{1}-n_{2}$,
would be of the most interest among them.
As we have presented in Ref. \cite{DKKN1}, 
the lattice SUSY invariant action with the asymmetric choice
can be constructed only on the integer sites,
so that one may expect 
the results without introducing any copy degrees of freedom.
On the other hand, it is important to recongize
that the superfield structure, 
particularly the structure of the vertex functions, are strongly 
related with the rotational symmetry in the configuration space. In particular, 
the relations (\ref{mass1})-(\ref{mass2}) 
and the cyclic permutation relations (\ref{int1})-(\ref{int2})
are satisfied only if one takes the symmetric choice of $a_{A}$ 
(\ref{a_symmetric_choice}).
In contrast,
if we take the asymmetric parameter choice, 
we do not have any cyclic permutational symmetry of the superfields 
at each vertex. In the configuration space, this corresponds to the fact that 
we do not have any non-trivial rotational symmetry if we take $a=0$.
The lack of the rotational symmetry in the case of $a=0$ makes
the perturbative calculations much more complicated and non-trivial 
even in terms of the superfields. Actually each three point 
vertex would give rise to six independent terms,
instead of two in the case of the symmetric choice. 
Accordingly, their $N$ dependences could get less transparent. 
The perturbative study of the asymmetric choice 
therefore should require more
extensive and careful calculations and observations.
We will keep this subject for future study 
and the result will be given elsewhere.
We should also note that 
addressing the rotational symmetry of this formulation 
may involve how to define 
the rotational symmetry itself 
within the context of non-commutative superspace framework.
We observe that  the ``twisted" deformation of Lorentz symmetry which has 
been proposed in the context of the Moyal non-commutative spacetime
in Ref. \cite{Chaichian} may play an important role also in 
the non-commutative lattice SUSY formulation.

Recently, the Leibniz rule on the lattice was investigated from 
an axiomatic point of view, 
and its physical relevance 
with the infinite flavor d.o.f. was pointed out
\cite{Kato-Sakamoto-So}.
It may be worthwhile to look at the 
non-commutative superspace formulation presented in this paper
from that point of view.

\section*{Acknowledgments}

The author would like to thank A. D'Adda, I. Kanamori and N. Kawamoto 
for useful discussions and comments.
This work is supported by Department of Energy US Government, 
Grant No. FG02-91ER 40661.


\end{document}